\documentclass[preprint,aps,prc,superscriptaddress,showpacs,floatfix]{revtex4}
\usepackage{graphicx}

\begin{document}

\title{Eta absorption by mesons}
\author{W. Liu and C. M. Ko}
\affiliation{Cyclotron Institute and Physics Department, Texas A$\&$M 
University, College Station, Texas 77843-3366}
\author{L. W. Chen}
\affiliation{Institute of Theoretical Physics, Shanghai Jiao Tong 
University, Shanghai 200240, China}

\begin{abstract}
Using the $[SU(3)_{\mathrm{L}} \times SU(3)_{\mathrm{R}}]_{\mathrm{global}%
}\times [SU(3)_V]_{\mathrm{local}}$ chiral Lagrangian with hidden local
symmetry, we evaluate the cross sections for the absorption of eta meson ($%
\eta$) by pion ($\pi$), rho ($\rho$), omega ($\omega$), kaon ($K$), and kaon
star ($K^*$) in the tree-level approximation. With empirical
masses and coupling constants as well as reasonable values for the 
cutoff parameter in the form factors at interaction vertices, we 
find that most cross sections are less than 1 mb, except the reactions 
$\rho\eta\to K\bar K^*(\bar KK^*)$, $\omega\eta\to K\bar K^*(\bar KK^*)$, 
$K^*\eta\to\rho K$, and $K^*\eta\to\omega K$, which are a few mb, and the
reactions $\pi\eta\to K\bar K$ and $K\eta\to\pi K$, which are more than 
10 mb. Including these reactions in a kinetic model based on a schematic 
hydrodynamic description of relativistic heavy ion collisions, we find 
that the abundance of eta mesons likely reaches chemical equilibrium
with other hadrons in nuclear collisions at the Relativistic Heavy Ion 
Collider.
\end{abstract}

\maketitle

\section{introduction}

Besides the intrinsic interest of studying eta meson production in heavy ion
collisions \cite{eta}, knowledge of the eta meson dynamics in the hot dense
matter formed in these collisions is also important for understanding other
observables measured in experiments. For example, to extract information on
the rho meson in-medium properties from the dilepton spectrum it emits
requires reliable knowledge on the number of eta mesons produced in heavy
ion collisions as eta mesons through their Dalitz decays contribute
significantly to low-mass dileptons measured in these collisions \cite%
{dilepton}. Also, the two-pion correlation function measured via the
Hanbury-Brown-Twiss interferometry is affected by pions from eta decays \cite%
{lin}. Understanding the eta meson dynamics is thus essential for extracting
the size of the pion emission source from the pion interferometry. In heavy
ion collisions at energies available from the Heavy Ion Syncrotron (SIS) at
the German Heavy Ion Research Center (GSI), whose dynamics is dominated by
baryons, the final eta number is mainly determined by its interaction with
the nucleon through the N(1535) resonance \cite{n1535}, which has a
branching ratio of about 35\% decaying into pion and nucleon. At higher
energies from the Super Proton Syncrotron (SPS) at the European Organization
for Nuclear Research (CERN), the matter becomes more dominated by mesons,
and the eta-meson interaction is thus important in determining the final eta
yield \cite{li}. For heavy ion collisions at the Relativistic Heavy Ion
Collider (RHIC) at Brookhaven National Laboratory (BNL), the final hadronic
matter is largely made of mesons, and it is even more crucial to have a good
knowledge on the cross sections for eta meson absorption by mesons.

Since there is no empirical information on these cross sections, theoretical
models are needed to evaluate their values. A possible model is the $[SU(3)_{%
\mathrm{L}} \times SU(3)_{\mathrm{R}}]_{\mathrm{global}} \times[SU(3)_V]_{%
\mathrm{local}}$ chiral Lagrangian with hidden local symmetry \cite{bando}.
This model has been used to study previously dilepton production \cite{song}
and more recently the interactions of phi mesons in hot hadronic matter \cite%
{koch}. In this paper, we shall use this model to evaluate the cross
sections for the absorption of $\eta$ meson by $\pi$ meson to the final
states of $K\bar{K}$, $K\bar K^*(\bar KK^*)$, $K^*\bar{K^*}$, $\rho\omega$,
and $\pi\rho$; by $\rho$ meson to the final states of $K\bar K$, 
$K\bar{K}^*(\bar{K}K^*)$, $K^*\bar K^*$, $\rho\rho$, $\pi\omega$, and 
$\pi\pi$; by $\omega$ meson to the final states of $K\bar K$, 
$K\bar K^*(\bar KK^*)$, $K^*\bar K^*$, and $\pi\rho$; and by $K$ and $K^*$ 
to the final states of $K(K^*)\pi$, $K(K^*)\rho$, and $K(K^*)\omega$. 
In evaluating the cross sections for these reactions, we do not include 
the effects due to scalar mesons $a(980)$ and $a(1450)$ as studies based 
on the non-linear chiral Lagrangian have shown that contributions from 
these resonances are unimportant compared to those from the vector meson 
exchanges and contact interactions \cite{black}. These cross sections 
are then used in a kinetic model to study their effects on eta meson 
abundance in heavy ion collisions at RHIC.

This paper is organized as follows. In Section \ref{hidden}, we give a brief
description of the $[SU(3)_{\mathrm{L}} \times SU(3)_{\mathrm{R}}]_{\mathrm{%
global}} \times[SU(3)_V]_{\mathrm{local}}$ chiral Lagrangian with hidden
local symmetry. The interaction Lagrangians that are relevant for 
describing the absorption of $\eta$ meson by $\pi$, $\rho$, $\omega$, $K$, 
and $K^*$ are derived in Section \ref{lagrangian}. The cross sections 
for these reactions are then evaluated in Section \ref{cs}. In Section 
\ref{heavyion}, the time evolution of $\eta$ meson abundance in 
relativistic heavy ion collisions is studied in a schematic model. 
Finally, a summary is given in Section \ref{summary}. Explicit 
expressions for the amplitudes for $\eta$ meson absorption are given in 
Appendices \ref{pion}-\ref{kstar}.

\section{chiral Lagrangian with hidden local symmetry}

\label{hidden}

Based on the hidden local gauge symmetry, Bando \cite{bando} has
constructed an effective theory for hadrons interacting at low energies. In
this approach, vector mesons are introduced as the gauge bosons of the
hidden local symmetry of the nonlinear chiral Lagrangian. For hadrons in the
SU(3) multiplets, which are relevant for present study, it is constructed
with two SU(3)-matrix valued variables $\xi_L(x)$ and $\xi_R(x)$ that
transform as $\xi_{L,R}(x) \rightarrow \xi^{\prime}_{L,R}(x)=h(x) \xi_{L,R}~
g^\dagger_{L,R}$ under $h(x)\in[$SU(3)$_V]_{\mathrm{local}}$ and $g_{L,R}\in[%
$SU(3)$_{L,R}]_{\mathrm{global}}$. The resulting chirally invariant
Lagrangian is 
\begin{eqnarray}
\mathcal{L} & = & \mathcal{L}_A + a\mathcal{L}_V-\frac{1}{2} \left\langle
F_{\mu\nu}F^{\mu\nu}\right\rangle,
\end{eqnarray}
with 
\begin{eqnarray}
\mathcal{L}_A & = & -\frac{1}{4}f^{2}_{\pi}\left\langle
(D_\mu\xi_L\cdot\xi_L^\dagger-D_\mu\xi_R\cdot\xi_R^\dagger)^2 \right\rangle,
\nonumber \\
\mathcal{L}_V & = & -\frac{1}{4}f^{2}_{\pi}\left\langle
(D_\mu\xi_L\cdot\xi_L^\dagger+D_\mu\xi_R\cdot\xi_R^\dagger)^2 \right\rangle,
\nonumber \\
F_{\mu\nu}&=&\partial_\mu V_\nu -\partial_\nu V_\mu -ig[V_\mu ,V_\nu].
\end{eqnarray}
In the above, $f_\pi$ is the pion decay constant, the symbol $%
\langle\cdot\cdot\rangle$ denotes the trace of $3\times 3$ matrix, and $%
D_\mu=\partial\mu-igV_\mu$ is the covariant derivative, with the dynamical
gauge bosons $V_\mu$ of the hidden local symmetry identified with the nonet
of vector mesons, i.e., 
\begin{eqnarray}  \label{vector}
V_\mu = \frac{1}{\sqrt{2}}\left(%
\begin{array}{ccc}
\frac{1}{\sqrt{2}}\rho_\mu^0 +\frac{1}{\sqrt{2}}\omega_\mu & \rho_\mu^+ & 
K_\mu^{*+} \\ 
\rho_\mu^- & -\frac{1}{\sqrt{2}}\rho_\mu^0 + \frac{1}{\sqrt{2}}\omega_\mu & 
K_\mu^{*0} \\ 
K_\mu^{*-} & \bar{K}_\mu^{*0} & \phi_\mu%
\end{array}%
\right).
\end{eqnarray}
The above effective Lagrangian with $a=2$ is known to give the universality
of vector meson coupling, the Kawarabayashi-Suzuki-Riazuddin-Fayyazuddin
(KSRF) relations \cite{ksfr}, and the vector meson dominance of the 
pseudoscalar meson electromagnetic form factor \cite{sakurai}.

Fixing the $[{\mbox SU(3)}_V]_{\mathrm{local}}$ gauge by 
\begin{eqnarray}
\xi_L^{\dagger} = \xi_{R}\equiv\xi = \exp(i\Phi /f_\pi),
\end{eqnarray}
where $\Phi$ is the nonet of pseudoscalar Goldstone bosons 
\begin{eqnarray}  \label{pseudoscalar}
\Phi = \frac{1}{\sqrt{2}}\left(%
\begin{array}{ccc}
\frac{1}{\sqrt{2}}\pi^0+\frac{1}{\sqrt{3}}\eta
+\frac{1}{\sqrt{6}}\eta^\prime & \pi^+ & K^{+} \\ 
\pi^- & -\frac{1}{\sqrt{2}}\pi^0+\frac{1}{\sqrt{3}}\eta
+\frac{1}{\sqrt{6}}\eta^\prime & K^{0} \\ 
K^{-} & \bar{K}^{0} & -\frac{1}{\sqrt{3}}\eta+\sqrt\frac{2}{3}\eta^\prime
\end{array}%
\right),
\end{eqnarray}
leads to the usual lowest-order chiral Lagrangian 
\begin{eqnarray}  \label{pseudo}
\mathcal{L}_A=-\frac{f_{\pi}^{2}}{4}\left\langle(\partial_{\mu}\xi\cdot
\xi^{\dagger}-\partial_{\mu}\xi^{\dagger}\cdot\xi)^2\right\rangle =\frac{%
f_\pi^2}{4}\left\langle\partial_{\mu}U\partial^{\mu}
U^{\dagger}\right\rangle,
\end{eqnarray}
with $U=\xi^2$. In obtaining Eq.(\ref{pseudoscalar}), we have used 
the empirical mixing angle $\theta\approx -20^\circ$ \cite{bramon74,particle}
between the octet $\eta_8$ and the singlet $\eta_1$ to obtain
the physical $\eta$ and $\eta^\prime$ via
$\eta\approx (2\sqrt{2}\eta_8+\eta_1)/3$ and
$\eta^\prime\approx (-\eta_8+2\sqrt{2}\eta_1)/3$.

The SU(3) symmetry breaking effects are taken into account by introducing in
the Lagrangian the mass term 
\begin{eqnarray}
\mathcal{L}_{SB}=\frac{1}{4}f_{\pi}^2\left\langle\xi_L\mathcal{M}%
\xi_R^{\dagger} +\xi_R\mathcal{M}\xi_L^{\dagger}\right\rangle,
\end{eqnarray}
with the mass matrix $\mathcal{M}$ given by 
\begin{eqnarray}
\mathcal{M}=\mathrm{diag}(m_{\pi}^2, m_{\pi}^2, 2m_K^2-m_{\pi}^2).
\label{mass}
\end{eqnarray}

Also, $\mathcal{L}_A$ and $\mathcal{L}_V$ are modified as follows \cite%
{bramon}: 
\begin{eqnarray}\label{breaking}
\mathcal{L}_A +\Delta\mathcal{L}_A &=& -\frac{1}{4}f_\pi^2\left\langle(D_\mu
\xi_L \cdot\xi_L^\dagger-D_\mu \xi_R \cdot\xi_R^\dagger)^2 (1+\xi_L
\epsilon_A \xi_R^\dagger +\xi_R \epsilon_A \xi_L^\dagger)\right\rangle, 
\nonumber  \label{la} \\
\mathcal{L}_V +\Delta\mathcal{L}_V &=& -\frac{1}{4}f_\pi^2\left\langle(D_\mu
\xi_L \cdot\xi_L^\dagger +D_\mu \xi_R \cdot\xi_R^\dagger)^2 (1+\xi_L
\epsilon_V \xi_R^\dagger + \xi_R \epsilon_V \xi_L^\dagger)\right\rangle,
\end{eqnarray}
with $\epsilon_{A(V)}= \mathrm{diag}(0,0,c_{A(V)}$), where $c_{A(V)}$ are
real parameters to be determined by empirical hadron masses and decay
constants.

Expanding the pseudoscalar fields in the first equation of Eq.(\ref{breaking})
shows that the kinetic terms for $K$ and $\eta$ are renormalized by the 
symmetry breaking term. The proper kinetic terms can be recovered by 
rescaling the $K$ and $\eta$ fields according to \cite{bramon} 
\begin{eqnarray}\label{scaling}
\sqrt{1+c_A}K\rightarrow K \ \ {\mbox{and}} \ \ \sqrt{1+\frac{2}{3}c_A}%
\eta\rightarrow \eta.
\end{eqnarray}
As a result, the kaon and eta decay constants are related to the pion decay
constant by 
\begin{eqnarray}\label{fk}
f_K =\sqrt{1+c_A}f_\pi \ \ {\mbox{and}} \ \ f_\eta =\sqrt{1+\frac{2}{3}c_A}
f_\pi,
\end{eqnarray}
and masses of vector mesons become different, i.e., 
\begin{eqnarray}\label{KSFR}
m_\rho^2 =m_\omega^2 =\frac{m_{K^*}^2}{1+c_V}=\frac{m_\phi^2}{1+2c_V}
=af_\pi^2g^2.
\end{eqnarray}

Expanding the Lagrangian up to four meson fields leads to interaction
Lagrangians of the VPP, VVV, VVPP, and PPPP types, i.e., 
\begin{eqnarray}
\mathcal{L}_{VPP}&=&-i\frac{a}{2}g\left\langle \left\{[\Phi ,\partial_\mu
\Phi ],V^\mu\right\} (1+2\epsilon_V )\right\rangle,  \nonumber \\
\mathcal{L}_{VVV}&=&ig\left\langle(\partial_\mu V_\nu-\partial_\nu V_\mu)
[V^\mu, V^\nu]\right\rangle,  \nonumber \\
\mathcal{L}_{VVPP}&=&-ag^2\left\langle V_\mu V^\mu (\epsilon_V \Phi^2 +2\Phi
\epsilon_V \Phi+\Phi^2 \epsilon_V )\right\rangle,  \nonumber \\
\mathcal{L}_{PPPP}&=&\frac{2}{3}\frac{1}{f_\pi^2} \left\langle\partial_\mu
\Phi \Phi\partial^\mu\Phi \Phi -\partial_\mu \Phi\partial^\mu \Phi
\Phi^2\right\rangle +\frac{1}{3f_{\pi}^2}\left\langle\mathcal{M}%
\Phi^4\right\rangle  \nonumber \\
&&-\frac{1}{f_\pi^2}\left\langle\partial_\mu \Phi \partial^\mu \Phi (2\Phi
\epsilon_A \Phi +\Phi^2 \epsilon_A +\epsilon_A \Phi^2) \right\rangle .
\label{interaction}
\end{eqnarray}

In the chiral Lagrangian with hidden local symmetry, vector mesons have also
been included through the anomalous interaction of the VVP type in order to
take into account the breaking of local chiral symmetry \cite{fujiwara}.
Including further the flavor breaking through a term $\xi_L\epsilon_{\mathrm{%
wz}}\xi^\dagger_R+\xi_R\epsilon_{\mathrm{wz}}\xi^\dagger_L$ with $\epsilon_{%
\mathrm{wz}}=\mathrm{diag}(0,0,c_{\mathrm{wz}})$ \cite{bramon}, the total
anomalous interaction Lagrangian then reads as 
\begin{eqnarray}  \label{anomalous}
\mathcal{L}_{VVP}+\Delta\mathcal{L}_{VVP}=2g_{VVP}\epsilon^{\mu\nu\lambda%
\sigma} \langle\partial_\mu V_\nu(1+2\epsilon_{\mathrm{wz}})
\partial_\lambda V_\sigma \Phi\rangle.
\end{eqnarray}

\section{Interaction Lagrangians}

\label{lagrangian}

Inserting Eqs.(\ref{vector}) and (\ref{pseudoscalar}) in Eqs.(\ref%
{interaction}) and (\ref{anomalous}) and rescaling the $K$ and $\eta$
meson fields according to Eq.(\ref{scaling}), we obtain following interaction
Lagrangian densities that are relevant to $\eta$ meson absorption by $\pi$, $%
\rho$, $\omega$, $K$, and $K^*$: 
\begin{eqnarray}\label{intlag}
\mathcal{L}_{\rho\pi\pi}&=&\frac{ag}{2}\vec\rho^\mu\cdot(\vec\pi\times
\partial_\mu\vec\pi),  \nonumber \\
\mathcal{L}_{\rho KK}&=& i\frac{ag}{4}\frac{1}{1+c_A} (\bar{K}\vec{\tau}%
\partial_\mu K- \partial_\mu \bar{K}\vec{\tau}K) \cdot\vec{\rho}^\mu, 
\nonumber \\
\mathcal{L}_{\omega KK}&=& i\frac{ag}{4}\frac{1}{1+c_A} (\bar{K}\partial_\mu
K- \partial_\mu \bar{K}K)\omega^\mu,  \nonumber \\
\mathcal{L}_{K^*K\pi}&=&i\frac{ag}{4}\frac{1+c_V}{\sqrt{1+c_A}}\bar{K}_\mu^* 
\vec{\tau}\cdot(K\partial^\mu \vec{\pi}-\partial^\mu K\vec{\pi}) \mathrm{%
+H.c.},  \nonumber \\
\mathcal{L}_{K^*K\eta}&=&i\frac{ag}{\sqrt{6}}\frac{1+c_V}{\sqrt{(1+c_A) (1+%
\frac{2}{3}c_A)}}\bar{K}_\mu^* (K\partial^\mu \eta -\partial^\mu
K\eta)+\mathrm{H.c.},  \nonumber \\
\mathcal{L}_{\rho\rho\rho}&=&g\partial_\mu\vec{\rho}_\nu \cdot(\vec{\rho}%
^\mu\times\vec{\rho}^\nu),  \nonumber \\
\mathcal{L}_{\rho K^*K^*}&=&i\frac{g}{2}[(\partial_\mu\bar K^{*\nu} \vec\tau
K^*_\nu-\bar K^{*\nu}\vec\tau\partial_\mu K^*_\nu) \cdot\vec\rho^\mu+(\bar
K^{*\nu}\vec\tau\cdot\partial_\mu \vec \rho_\nu -\partial_\mu\bar
K^{*\nu}\vec\tau\cdot \vec\rho_\nu)K^{*\mu}  \nonumber \\
&&+\bar K^{*\mu}(\vec\tau\cdot\vec\rho^\nu\partial_\mu K^*_\nu -\vec\tau\cdot\partial_\mu
\vec\rho^\nu K^*_\nu)],  \nonumber \\
\mathcal{L}_{\omega K^*K^*}&=&i\frac{g}{2}[(\partial_\mu\bar K^{*\nu}
K^*_\nu -\bar K^{*\nu}\partial_\mu K^*_\nu)\omega^\mu +(\bar
K^{*\nu}\partial_\mu \omega_\nu -\partial_\mu\bar K^{*\nu}\omega_\nu)K^{*\mu}
\nonumber \\
&&+\bar K^{*\mu}(\omega^\nu\partial_\mu K^*_\nu -\partial_\mu \omega^\nu
K^*_\nu)],  \nonumber \\
\mathcal{L}_{\rho K^*K\eta}&=&\frac{ag^2}{2\sqrt{6}}\frac{c_V}{\sqrt{%
(1+c_A) (1+\frac{2}{3}c_A)}}(\bar{K}\vec{\tau}K^*_\mu\cdot\vec\rho^\mu\eta 
+\bar{K}^*_\mu\vec{\tau}K\cdot \vec{\rho}^\mu\eta),  \nonumber \\
\mathcal{L}_{\omega K^*K\eta}&=&\frac{ag^2}{2\sqrt{6}}\frac{c_V}{\sqrt{%
(1+c_A) (1+\frac{2}{3}c_A)}}(\bar{K}K^*_\mu\omega^\mu \eta
+\bar{K}^*_\mu K\omega^\mu\eta),  \nonumber \\
\mathcal{L}_{\pi\eta KK} &=& \frac{1}{3\sqrt{6}f_\pi^2} \frac{1}{(1+c_A) 
\sqrt{1+\frac{2}{3}c_A}}\left[\left(1+\frac{3}{2}c_A \right)(\bar K
\vec\tau\partial_\mu K \cdot\partial^\mu \vec{\pi}\eta
+\partial_\mu\bar K\vec\tau K \cdot\partial^\mu\vec\pi\eta)\right. \nonumber \\
&&+\bar K\vec\tau\partial_\mu K \cdot \vec\pi\partial^\mu\eta
+\partial_\mu\bar K\vec\tau K \cdot\vec\pi\partial^\mu\eta 
-\left(2+3c_A\right)\bar{K}\vec{\tau}K\cdot\partial_\mu \vec{\pi}
\partial^\mu \eta \nonumber\\
&&-\left.2\partial_\mu\bar K\vec\tau\partial^\mu K\cdot\vec\pi\eta 
+m_\pi^2\bar{K} \vec{\tau}K\cdot\vec{\pi}\eta\right].  \nonumber \\
\mathcal{L}_{\rho\rho\eta}&=&\frac{g_{\rho\rho\eta}}{\sqrt{6}}
\frac{1}{\sqrt{1+\frac{2}{3}c_A}}\epsilon^{\mu\nu\lambda\sigma}\partial_\mu 
\vec{\rho}_\nu\cdot\partial_\lambda\vec\rho_\sigma\eta,  \nonumber \\
\mathcal{L}_{\omega\omega\eta}&=&\frac{g_{\omega\omega\eta}}{\sqrt{6}}
\frac{1}{\sqrt{1+ \frac{2}{3}c_A}}\epsilon^{\mu\nu\lambda\sigma}
\partial_\mu\omega_\nu\partial_\lambda\omega_\sigma\eta,  \nonumber \\
\mathcal{L}_{\rho\omega\pi}&=&g_{\rho\omega\pi}
\epsilon^{\alpha\beta\lambda\sigma}\partial_\mu\vec\rho_\nu\partial_\lambda
\omega_\sigma\cdot\vec\pi,  \nonumber\\
\mathcal{L}_{\rho K^*K}&=&\frac{g_{\rho K^*K}}{2}\frac{1}{\sqrt{1+c_A}}
\epsilon^{\mu\nu\lambda\sigma}\left(\bar K\partial_\mu
{\vec \rho}_\nu\cdot{\vec\tau}\partial_\lambda K^*_\sigma
+\partial_\mu{\vec\rho}_\nu\cdot\partial_\lambda\bar K^*_\sigma{\vec\tau} 
K\right),  \nonumber \\
\mathcal{L}_{\omega K^*K}&=&\frac{g_{\omega K^*K}}{2}\frac{1}{\sqrt{1+c_A}}
\epsilon^{\mu\nu\lambda\sigma}\partial_\mu\omega_\nu \left(\bar
K\partial_\lambda K^*_\sigma +\partial_\lambda\bar K^*_\sigma K\right), 
\nonumber \\
\mathcal{L}_{K^*K^*\pi}&=&\frac{g_{K^*K^*\pi}}{2}(1+2c_{\mathrm{wz}})
\epsilon^{\mu\nu\lambda\sigma}\partial_\mu \bar K^*_\nu
\vec\tau\partial_\lambda K^*_\sigma\cdot\vec\pi,  \nonumber \\
\mathcal{L}_{K^*K^*\eta}&=&\frac{2g_{K^*K^*\eta}}{\sqrt{6}}
\frac{c_{\rm wz}}{\sqrt{1+\frac{2}{3}c_A}}
\epsilon^{\mu\nu\lambda\sigma} \partial_\mu\bar
K^*_\nu\partial_\lambda K^*_\sigma\eta.
\end{eqnarray}
In the above, $\vec{\tau}$ are Pauli matrices for isospin; and $\vec{\pi}$
and $\vec{\rho}$ denote the pion and rho meson isospin triplet,
respectively; and $K=(K^+ ,K^0)^T$ and $K^* =(K^{*+} ,K^{*0})^T$ denote the
pseudoscalar and vector strange meson isospin doublet, respectively.

For the parameters in the interaction Lagrangians given in 
Eq.(\ref{intlag}), we choose $a=2$ to recover the vector dominance model 
and the KSFR relation. Using the empirical value $f_\pi =92.4$ MeV 
for pion decay constant, $m_\rho=776$ MeV for rho meson mass, and 
$m_{K^*}=892$MeV for $K^*$ mass, the KSFR relation (Eq.(\ref{KSFR})) then 
gives a a universal vector coupling constant of $g=5.94$ and 
the parameter $c_V =0.32$. We note that a slightly larger value 
of $c_V=0.36$ was obtained in Ref.\cite{koch} by fitting the empirical 
phi meson mass $m_\phi=1020$ MeV. Since only the $K^*$ is needed 
in present study, we choose to have the correct $K^*$ instead
phi meson mass. From Eq.(\ref{fk}), the parameter $c_A=0.49$
is obtained from the empirical kaon decay constant $f_K=1.22 f_\pi$. 

It is interesting to know that above parameters lead to a width 
$\Gamma_\rho=g^2(m_\rho^2-4m_\pi^2)^{3/2}/(48\pi m_\rho^2)\sim 147$ MeV
for rho meson decay to two pions and a width
$\Gamma_{K^*}=[g^2(1+c_V)^2)/(1+c_A)]\{[m_{K^*}^2-(m_K+m_\pi)^2]
[m_{K^*}^2-(m_K-m_\pi)^2]\}^{3/2}/(16\pi m_{K^*}^5)\sim 49$ MeV
for $K^*$ decay to $K\pi$, which are close to the empirical values 
of $\Gamma_\rho=150$ MeV and $\Gamma_{K^*}=51$ MeV. 

For the anomalous coupling constants $g_{VVP}$, they are
related to the universal vector coupling $g$ and pseudoscalar decay 
constant by \cite{ksfr}, 
\begin{eqnarray}
g_{VVP}=\frac{3g^2}{8\pi^2 f_P},~~~~~~P=\pi,~\eta,~K
\end{eqnarray}
and from the ratio between experimental decay widths of 
$K^{*0}\to K^0\gamma$ and $K^{*\pm}\to K^\pm\gamma$, the
parameter $c_{\mathrm{wz}}=-0.1$ has been determined \cite{bramon}.

\section{eta absorption by mesons}

\label{cs}

Since the temperature of the hot hadronic matter produced at RHIC is above $%
T=125$ MeV, it consists of not only the lightest $\pi$ mesons but also
heavier $\rho$ and $\omega$ mesons as well as the strange mesons $K$ and $K^*
$. Eta mesons can thus be absorbed by all these mesons. 

\subsection{Born diagrams}

\begin{figure}[ht]
\includegraphics[width=4.5in,height=4.5in,angle=0]{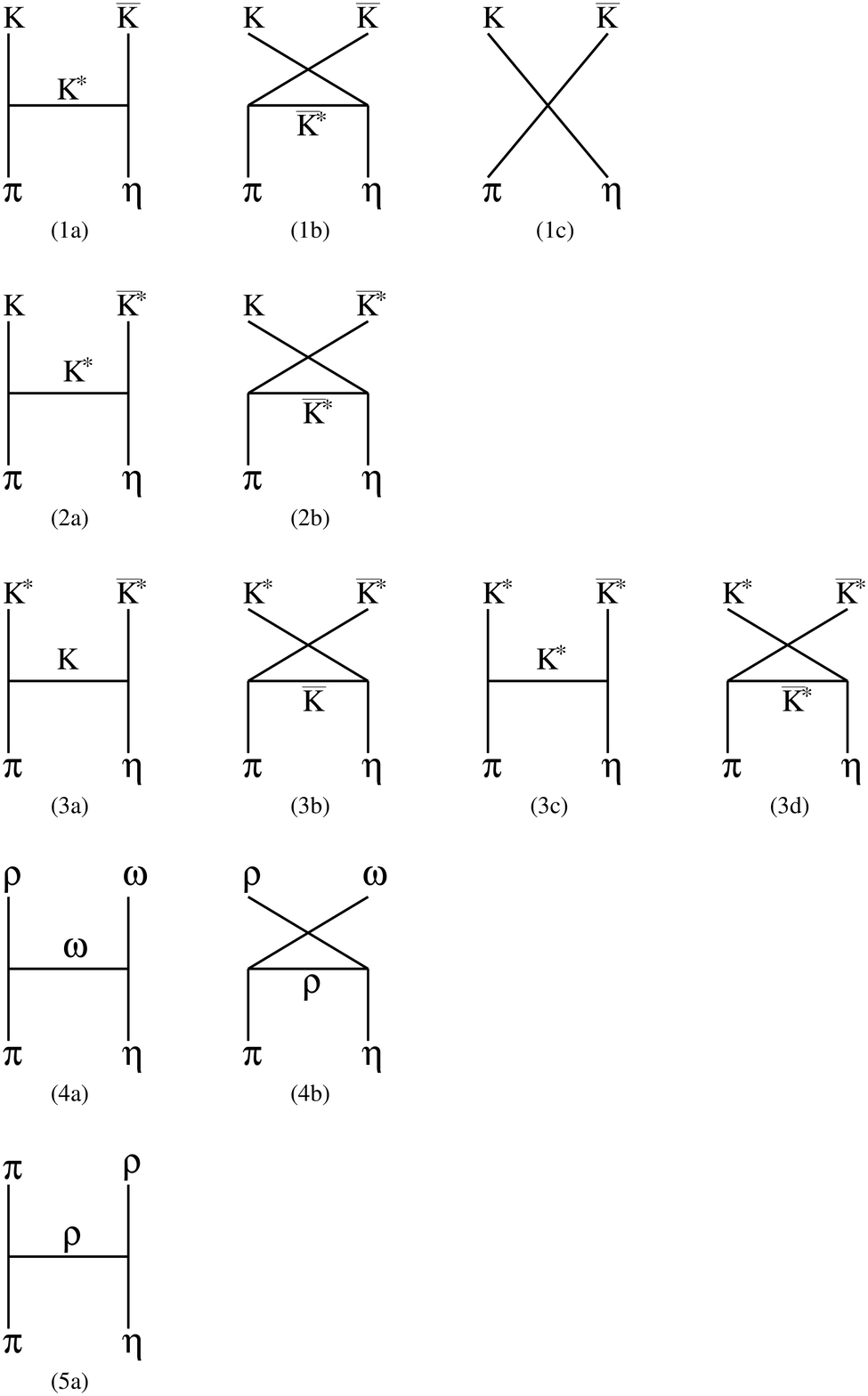} \vspace{0.5cm}
\caption{Diagrams for $\protect\eta$ absorption by $\protect\pi$ meson.}
\label{dpion}
\end{figure}

\begin{figure}[ht]
\includegraphics[width=4.5in,height=4.5in,angle=0]{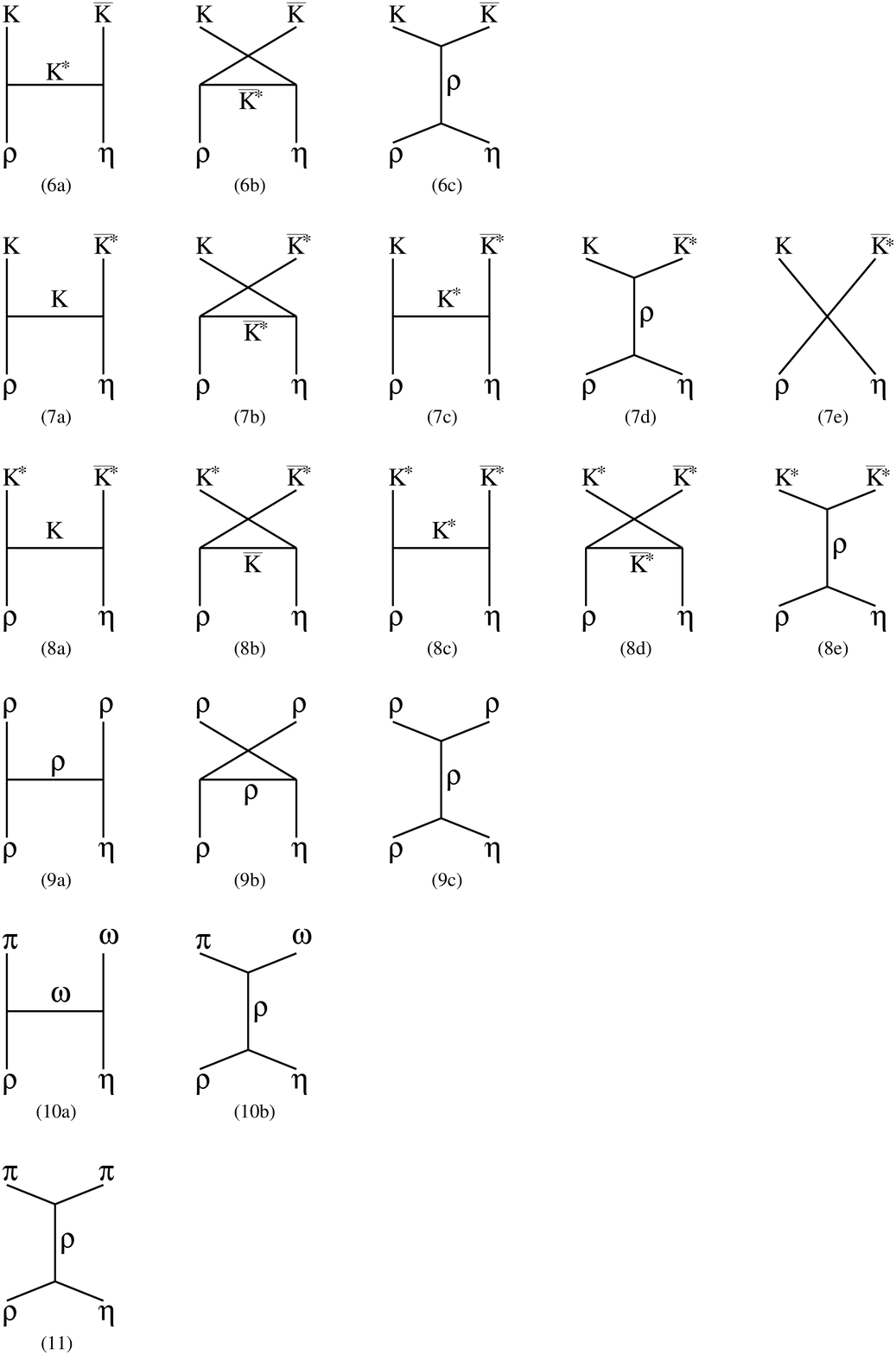} \vspace{0.5cm}
\caption{Diagrams for $\protect\eta$ absorption by $\protect\rho$ meson.}
\label{drho}
\end{figure}

\begin{figure}[ht]
\includegraphics[width=4.5in,height=4.5in,angle=0]{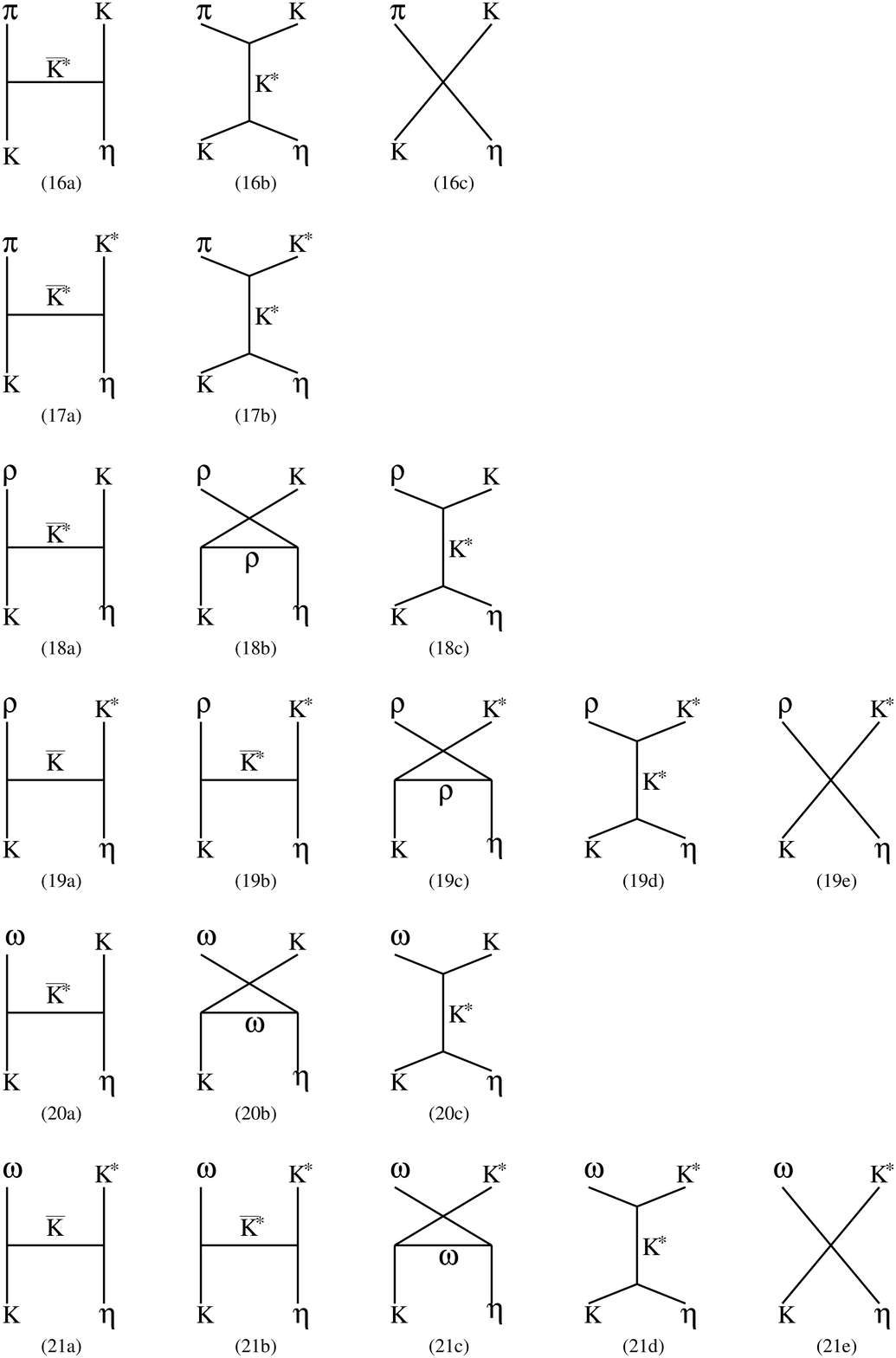} \vspace{0.5cm}
\caption{Diagrams for $\protect\eta$ absorption by $K$ meson.}
\label{dkaon}
\end{figure}

\begin{figure}[ht]
\includegraphics[width=4.5in,height=4.5in,angle=0]{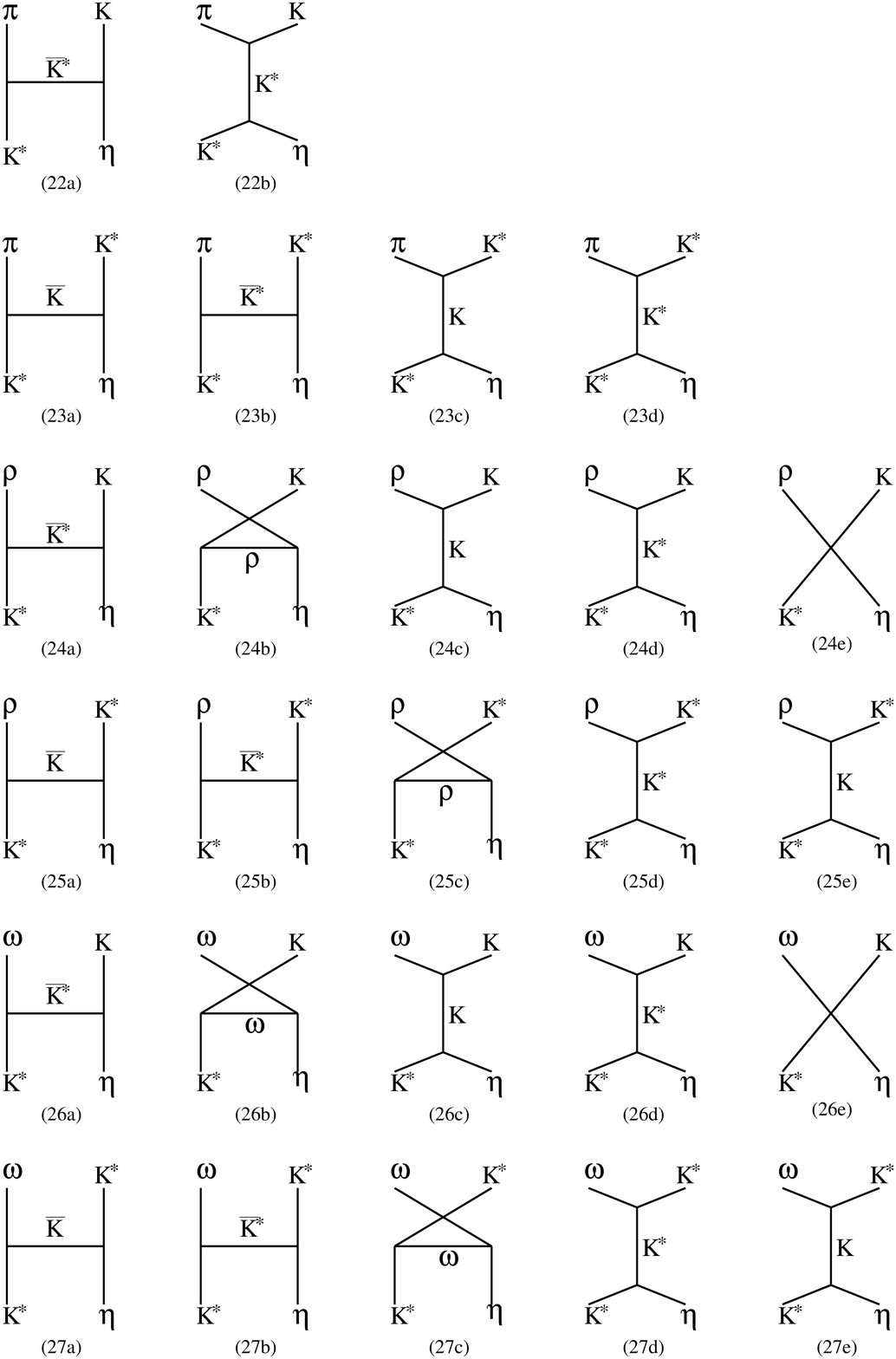} \vspace{0.5cm}
\caption{Diagrams for $\protect\eta$ absorption by $K^*$ meson.}
\label{dkstar}
\end{figure}

Diagrams for these reactions are shown in Fig. \ref{dpion} 
for $\eta$ absorption by $\pi$ to the the final states of 
$K\bar{K}$, $K\bar K^*(\bar KK^*)$, $K^*\bar{K^*}$,
$\rho\omega$, and $\pi\rho$; in Fig. \ref{drho} for $\eta$ absorption by 
$\rho$ meson to the final states $K\bar K$, $K\bar{K}^*(\bar{K}K^*)$, 
$K^*\bar K^*$, $\rho\rho$, $\pi\omega$, and $\pi\pi$; in Fig. \ref%
{domega} for $\eta$ absorption by $\omega$ meson to the final states $K\bar K$%
, $K\bar K^*$, $K^*\bar K^*$, and $\pi\rho$; and in Figs. \ref{dkaon} and 
\ref{dkstar} for $\eta$ absorption by $K$ and $K^*$ to the final states 
of $\pi K$, $\pi K^*$, $\rho K$, $\rho K^*$, $\omega K$, and $\omega K^*$.

\begin{figure}[ht]
\includegraphics[width=4.5in,height=3.0in,angle=0]{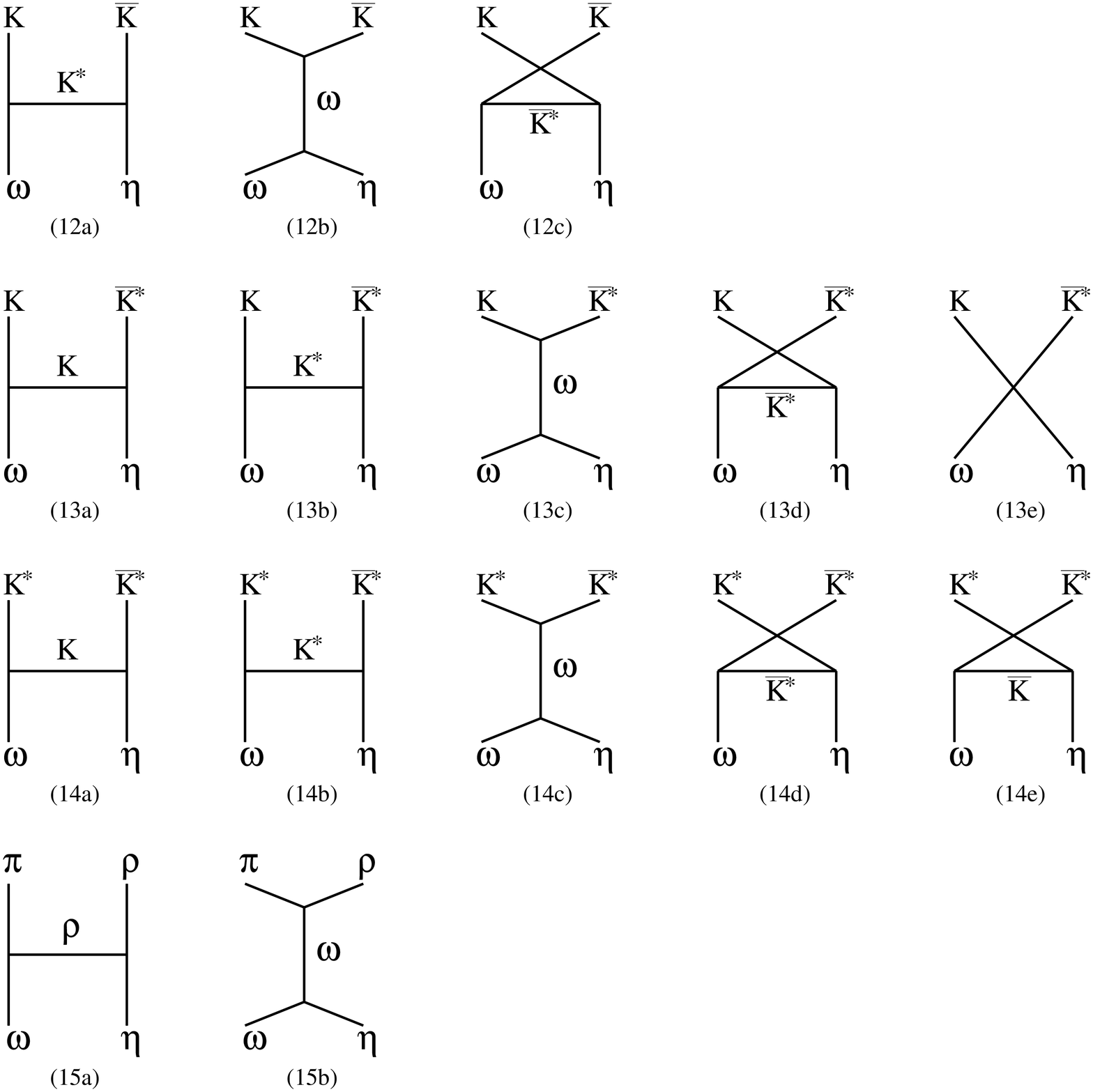} \vspace{%
0.5cm}
\caption{Diagrams for $\protect\eta$ absorption by $\protect\omega$ meson.}
\label{domega}
\end{figure}

Using the interaction Lagrangian densities given in Section \ref{lagrangian}%
, we have derived the amplitudes for all tree-level diagrams shown in Figs. %
\ref{dpion}-\ref{dkstar}. In general, the amplitude for a process $n$ is
given by 
\begin{eqnarray}
\mathcal{M}_n =A\left ( \sum_{i} \mathcal{M}_{n i}^{\lambda_k \cdots
\lambda_l} \right ) \epsilon_{k \lambda_k} \cdots \epsilon_{l \lambda_l},
\end{eqnarray}
where $i$ runs through all the subprocesses in each reaction, and $%
\epsilon_{j\lambda_j}$ denotes the polarization vector of external vector
meson $j$. The factor $A$ is either a matrix element $\tau^a_{ij}$ of the
Pauli matrices or the kronecker delta $\delta_{ab}$ or
the antisymmetric tensor $i\epsilon_{abc}$. It takes into
account the isospin states of the particles in a reaction, with $a$, $b$,
and $c$ denoting those of isospin triplet $\pi$ and $\rho$ meson, and $i$
and $j$ those of isospin doublet $K$ and $K^*$. Explicit expressions for
these amplitudes are given in Appendix \ref{pion} for $\eta$ absorption by $%
\pi$ meson, Appendix \ref{rho} for $\eta$ absorption by $\rho$ meson,
Appendix \ref{omega} for $\eta$ absorption by $\omega$ meson, Appendix \ref%
{kaon} for $\eta$ absorption by $K$ meson, and Appendix \ref{kstar} for $\eta
$ absorption by $K^*$ meson.

\subsection{form factors}

To obtain the full amplitudes for these reactions, one needs in principle to
carry out a coupled-channel calculation in order to avoid the violation of
unitarity. Such an approach is, however, beyond the scope of present work.
To prevent the artificial growth of the tree-level amplitudes with the
energy, we introduce instead form factors at interaction vertices and treat
their cutoff parameters as parameters. Specifically, the form factors are
taken to have the same forms as used previously in studying $J/\psi$
absorption \cite{jpsi1,jpsi2} and charmed meson scattering \cite{charm} by
mesons, charmed meson production from photon- and proton-proton reactions 
\cite{liu}, pentaquark baryon production from photon- and hadron-proton
reactions \cite{liu1}, and strangeness-exchange reactions between mesons and
baryons \cite{chli}. For three-point vertices, i.e, $\rho\pi\pi$, $\rho KK$, 
$\omega KK$, $K^*K\pi$, $K^*K\eta$, $\rho\rho\rho$, $\rho K^*K^*$, $\omega
K^*K^*$, $\rho\omega\pi$, $\rho\rho\eta$, $\omega\omega\eta$, $\rho K^*K$, 
$\omega K^*K$, $K^*K^*\pi$, and $K^*K^*\eta$ in the $t-$ and $u$-channel 
processes, the form factors are taken to have the form 
\begin{eqnarray}  \label{form1}
F_3(\mathbf{q}) =\frac{\Lambda^2}{\Lambda^2 +\mathbf{q}^2},
\end{eqnarray}
where $\mathbf{q}^2$, taken in the center of mass, is the squared three
momentum transfer. For four-point contact vertices, i.e., 
$\rho K^*K\eta$, $\omega K^*K\eta$, and $\pi\eta KK$, they are taken 
to have the form 
\begin{eqnarray}\label{form2}
F_4 =\left( \frac{\Lambda_1^2}{\Lambda_1^2 +\langle\mathbf{q}^2\rangle}%
\right) \left( \frac{\Lambda_2^2}{\Lambda_2^2 +\langle\mathbf{q}^2\rangle}%
\right),
\end{eqnarray}
where $\Lambda_1$ and $\Lambda_2$ are two different cutoff parameters at the
three-point vertices present in the processes with the same initial and
final particles, and $\langle\mathbf{q}^2\rangle$ is the average value of
the two squared three momenta at the three-point vertices. 

For three-point vertices in $s$-channel processes, we introduce, however, 
a covariant form factor, i.e., 
\begin{equation}
F_3^\prime(s) =\frac{\Lambda^2+m^2}{\Lambda^2+s},
\end{equation}
where $s$ is the square of the center-of-mass energy and $m$ is the mass
of the intermediate-state particle. This form factor ensures that it 
would not affect the predicted decay widths of $\rho$ and $K^*$,  
which are close to the empirical ones, as $s=m^2$ in these processes. 
We note that unrealistic large cross sections for $s-$channel processes 
at high energies can also be prevented by taking into account the 
vacuum widths for all channels that are open at a given $s$ as in 
Ref.\cite{koch} on phi meson interactions in hot hadronic matter.

As in previous studies on hadronic reactions, we use $\Lambda=1$ GeV 
in calculating the cross sections for eta absorption by mesons. Since 
the value for $\Lambda$ might depend on the reactions, we also study 
how our results are affected by changes in its value. 

\subsection{cross sections for eta absorption by mesons}

The isospin- and spin-averaged differential cross sections for above
reactions are given by 
\begin{eqnarray}  \label{cross}
\frac{d\sigma_n}{dt}=\frac{1}{64\pi sp_i^2N_IN_S}\overline{|\mathcal{M}_n|^2}%
,  \label{crosec}
\end{eqnarray}
where $\overline{|\mathcal{M}_n|^2}$ denotes the squared amplitude obtained
from summing over the isospins and spins of both initial and final
particles, and can be evaluated using the software package FORM \cite{form}.
The factors $N_I=(2I_1+1)(2I_2+1)$ and $N_S=(2S_1+1)(2S_2+1)$ in the
denominator are due to averaging over the isospins $I_1$ and $I_2$ as well
as the spins $S_1$ and $S_2$ of initial particles, while $p_i$ is their
3-momentum in the center-of-mass frame.

Integrating the four momentum transfer $t$ leads to the following total
cross sections: 
\begin{equation}  \label{tcross}
\sigma_n=\frac{1}{N_IN_S}\frac{p_f}{p_i}|M_n|^2,
\end{equation}
where $p_f$ is the 3-momentum of final particles in their center-of-mass
frame and $|M_n|^2$ is related to $\overline{|\mathcal{M}_n|^2}$ in Eq.(\ref
{cross}) by 
\begin{equation}  \label{matrix}
|M_n|^2=\frac{1}{64\pi^2 s}\int d\Omega\overline{|\mathcal{M}_n|^2}F^4,
\end{equation}
with $F$ denoting the appropriate form factors at interaction vertices.

\begin{figure}[ht]
\begin{minipage}{0.32 \textwidth}
\includegraphics[width=3.0in,height=2.5in,angle=270]{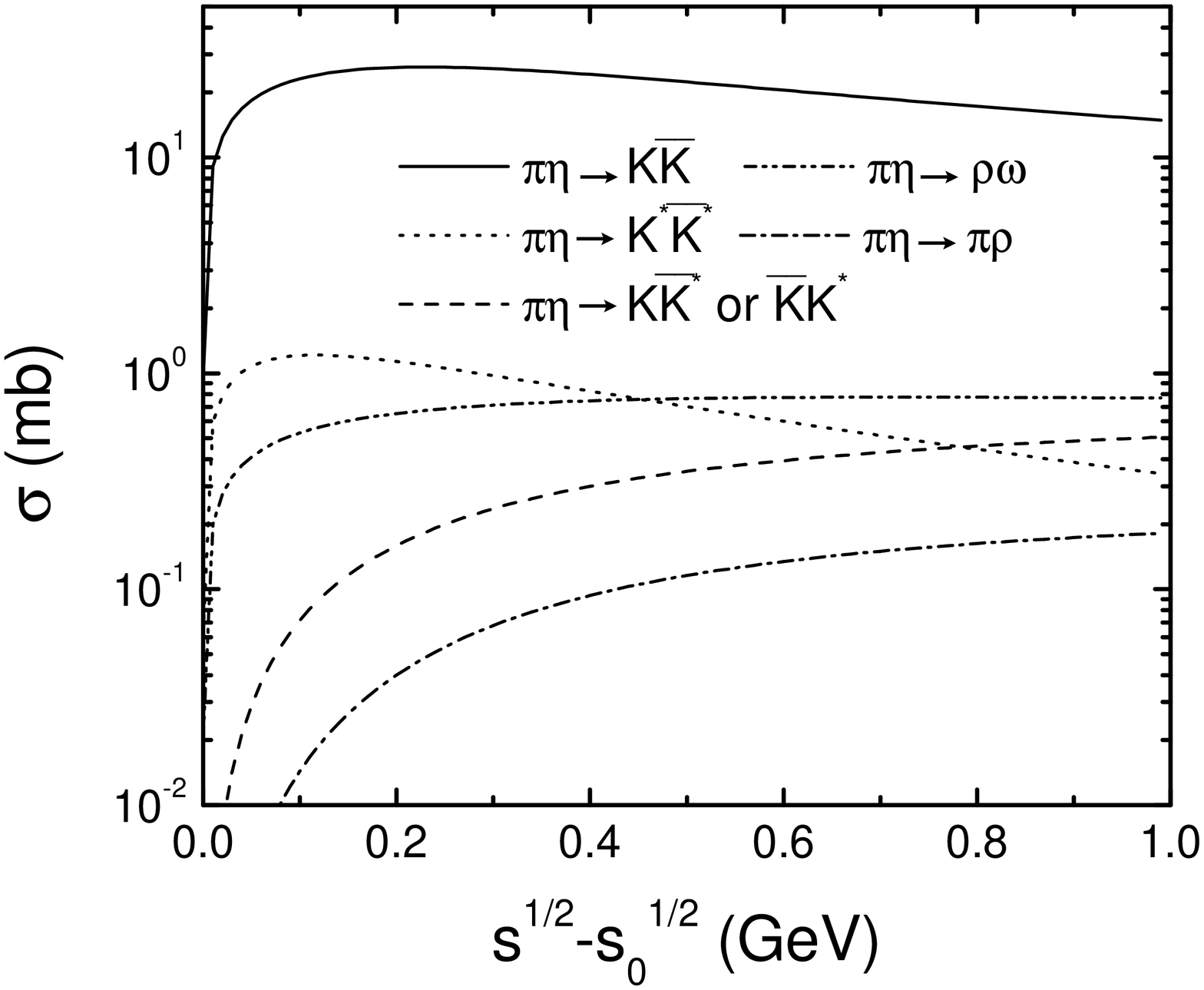}
\end{minipage}
\begin{minipage}{0.32 \textwidth}
\includegraphics[width=3.0in,height=2.5in,angle=270]{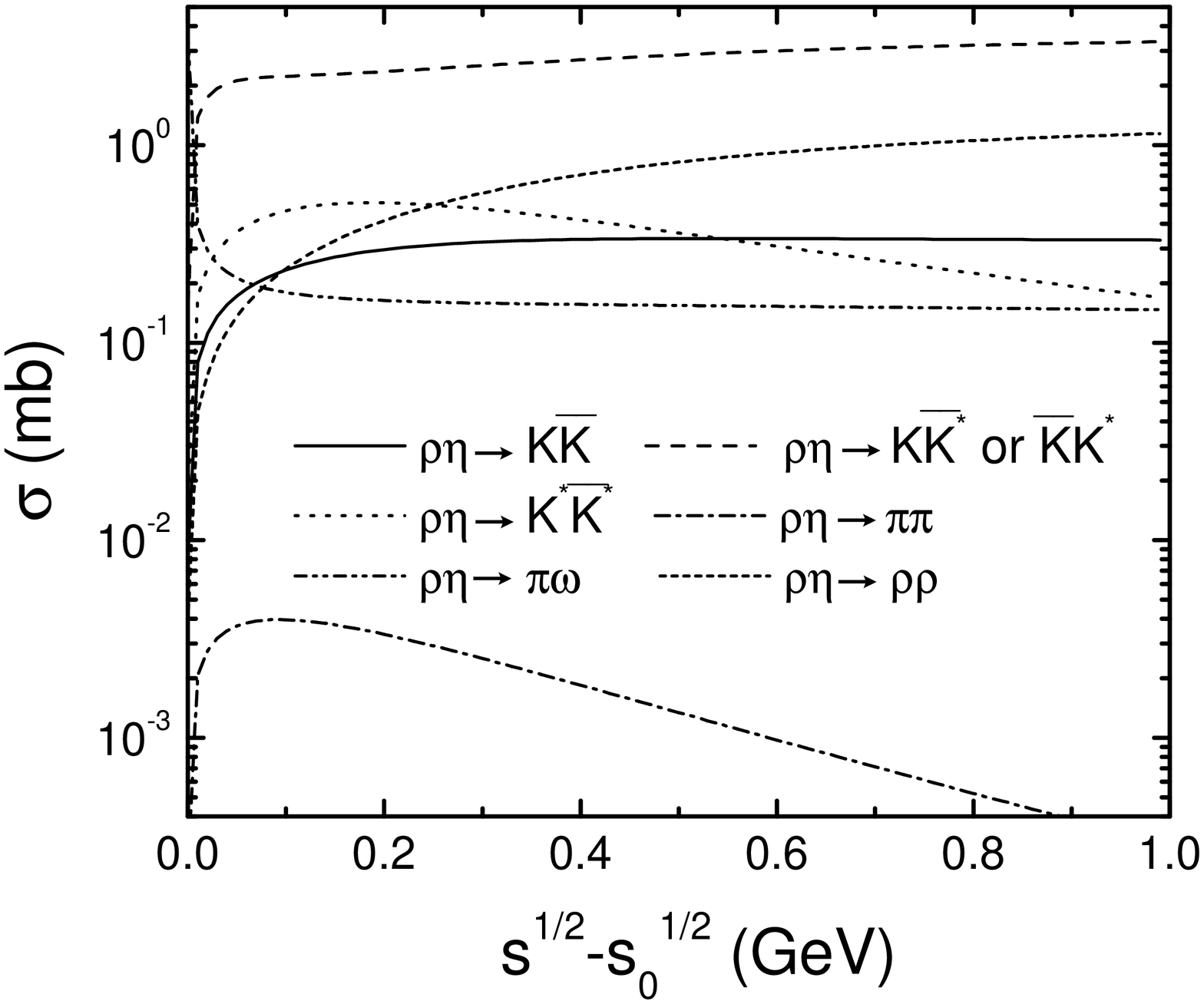}
\end{minipage}
\begin{minipage}{0.32 \textwidth}
\includegraphics[width=3.0in,height=2.5in,angle=270]{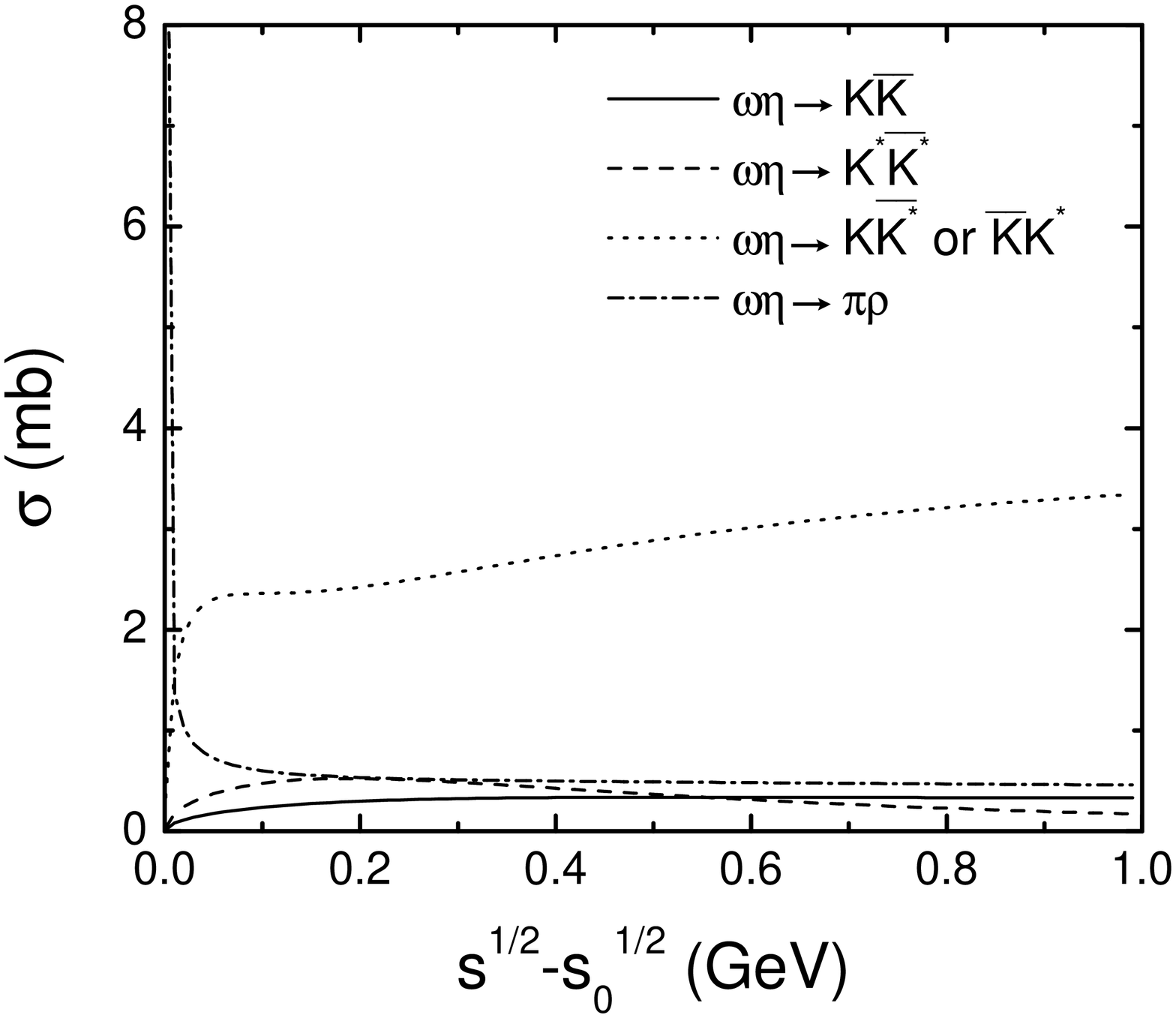}
\end{minipage}
\vspace{0.5cm}
\caption{Cross sections for eta absorption by $\protect\pi$ (left panel), $
\protect\rho$ (middle panel), and $\protect\omega$ (right panel) mesons as
functions of total center-of-mass energy $s^{1/2}$ above the threshold $%
s_0^{1/2}$ of a reaction.}
\label{cross1}
\end{figure}

\begin{figure}[ht]
\begin{minipage}{0.48 \textwidth}
\includegraphics[width=3.0in,height=3.0in,angle=270]{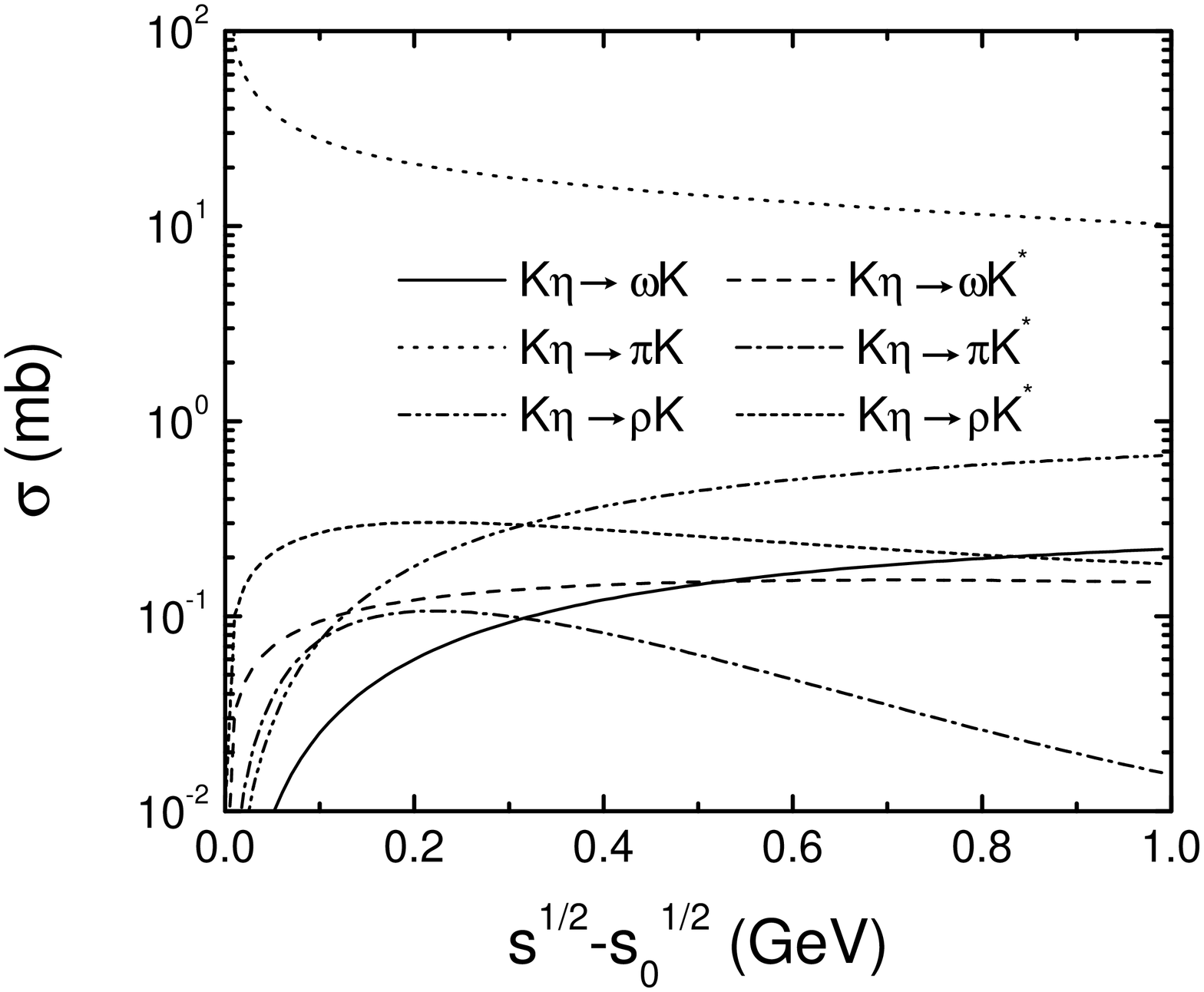}
\end{minipage}
\begin{minipage}{0.48 \textwidth}
\includegraphics[width=3.0in,height=3.0in,angle=270]{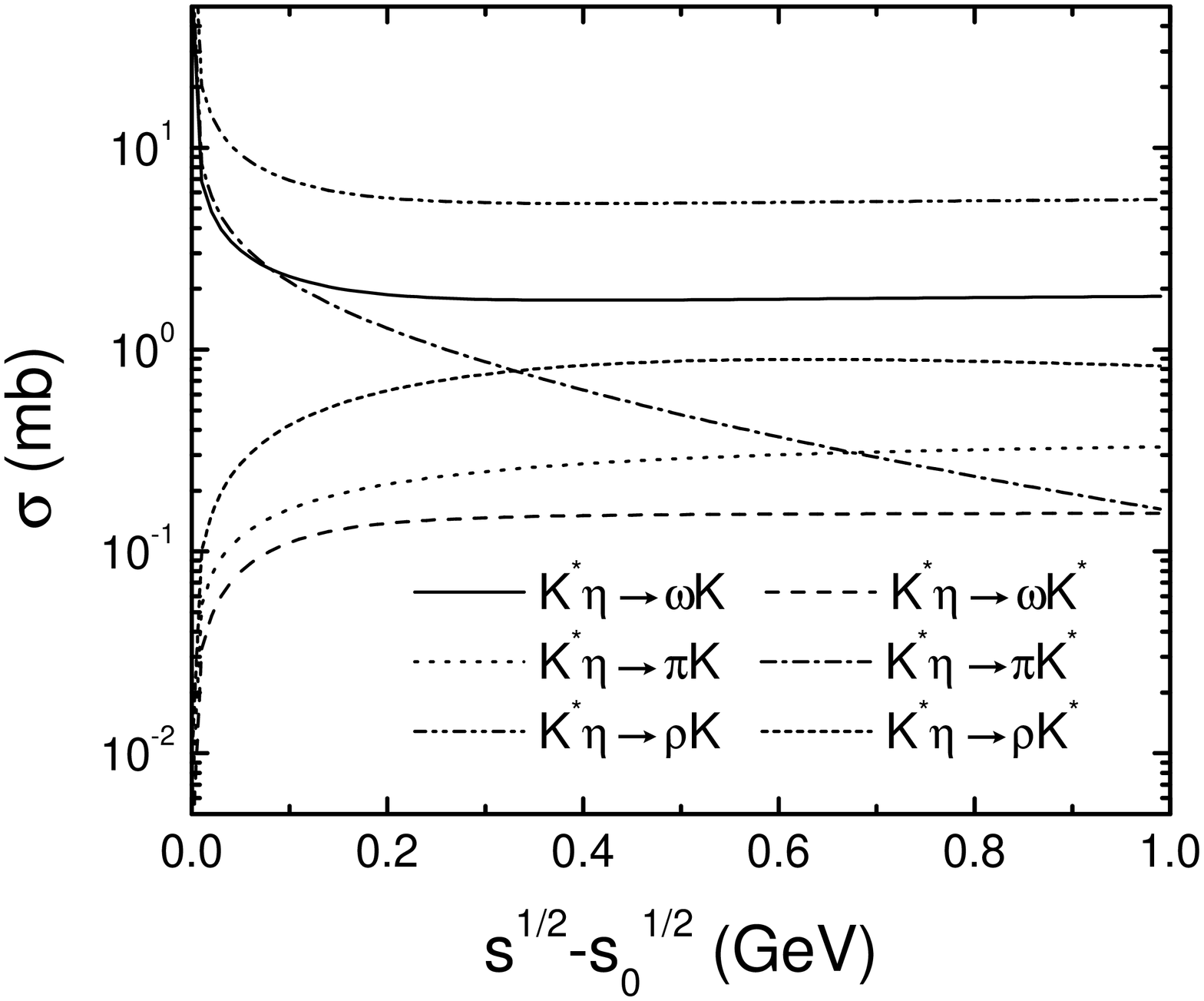}
\end{minipage}
\vspace{0.5cm}
\caption{Cross sections for eta absorption by $K$ (left panel) and $K^*$
(right panel) mesons as functions of total center-of-mass energy $s^{1/2}$
above the threshold $s_0^{1/2}$ of a reaction.}
\label{cross2}
\end{figure}

In Figs. \ref{cross1} and \ref{cross2}, we show, respectively, the cross
sections for $\eta$ absorption by $\pi$, $\rho$, and $\omega$ mesons and by $%
K$ and $K^*$ as functions of total center-of-mass energy $s^{1/2}$ above the
threshold $s_0^{1/2}$ of a reaction. Aside near the threshold of a reaction,
where the cross section can be very large or small depending on whether the
reaction is exothermic or endothermic, most cross sections are less than 1
mb, except the reactions $\rho\eta\to K\bar K^*(\bar KK^*)$, $\omega\eta\to
K\bar K^*(\bar KK^*)$, $K^*\eta\to\rho K$, and $K^*\eta\to\omega K$, which
are a few mb, and the reactions $\pi\eta\to K\bar K$ and $K\eta\to\pi K$, 
which are more than 10 mb. The large cross sections for the reactions 
$\pi\eta\to K\bar K$ and $K\eta\to\pi K$ are due to the presence of 
the $\pi\eta KK$ contact interaction in their amplitudes. Values of 
the cross sections depend, however, on the value of the cutoff parameter 
in the form factors. We find that increasing the value of $\Lambda$ 
to 2 GeV increases the cross sections by about a factor of two,
but decreasing its value to 0.5 GeV reduces the cross sections by about a
factor of four.

\begin{table}[tbp]
\caption{Values of parameters $a$, $b$, $c$, $d$, and $e$ in Eq.(\protect\ref%
{parameter}) for parameterizing the matrix elements defined in Eq.(\protect
\ref{matrix}).}%
\begin{tabular}{cccccc}
\hline\hline
Reactions & $\quad a$ & $\quad b$ & $c$ & $d$ & $e$ \\ \hline
$\pi \eta \rightarrow K\bar{K}$ & $16.047$ & $-468.921$ & $-3.867$ & $598.598
$ & $-3.215$ \\ 
$\pi \eta \rightarrow K\bar{K}^{\ast }(\bar{K}K^{\ast })$ & $5.813$ & $-4.232
$ & $-0.136$ & $-1.580$ & $-5.357$ \\ 
$\pi \eta \rightarrow K^{\ast }\bar{K^{\ast }}$ & $0.115$ & $7.261$ & $%
-13.200$ & $8.468$ & $-4.530$ \\ 
$\pi \eta \rightarrow \rho \omega $ & $1.539$ & $2.577$ & $-1.513$ & $0.555$
& $-15.636$ \\ 
$\pi \eta \rightarrow \pi \rho $ & $0.721$ & $-7.012$ & $-3.948$ & $6.293$ & 
$-4.323$ \\ \hline
$\rho \eta \rightarrow K\bar{K}$ & $2.706$ & $-7.532$ & $-5.657$ & $4.827$ & 
$-4.857$ \\ 
$\rho \eta \rightarrow K\bar{K}^{\ast }(\bar{K}K^{\ast })$ & $36.139$ & $%
20.087$ & $-18.884$ & $-17.458$ & $-2.338$ \\ 
$\rho \eta \rightarrow K^{\ast }\bar{K^{\ast }}$ & $0.181$ & $-4.174$ & $%
-19.494$ & $13.793$ & $-4.941$ \\ 
$\rho \eta \rightarrow \rho \rho $ & $14.265$ & $-2.296$ & $-2.812$ & $%
-10.394$ & $-2.812$ \\ 
$\rho \eta \rightarrow \pi \pi $ & $0.001$ & $-0.529$ & $-8.081$ & $0.529$ & 
$-7.476$ \\ 
$\rho \eta \rightarrow \pi \omega $ & $2.007$ & $-0.699$ & $-5.410$ & $-0.657
$ & $0.228$ \\ \hline
$\omega \eta \rightarrow K\bar{K}$ & $1.863$ & $-0.948$ & $-6.925$ & $-0.878$
& $0.056$ \\ 
$\omega \eta \rightarrow K\bar{K}^{\ast }(\bar{K}K^{\ast })$ & $12.179$ & $%
7.285$ & $-19.514$ & $-5.766$ & $-2.153$ \\ 
$\omega \eta \rightarrow K^{\ast }\bar{K^{\ast }}$ & $0.060$ & $-1.434$ & $%
-19.684$ & $4.669$ & $-4.948$ \\ 
$\omega \eta \rightarrow \pi \rho $ & $7.784$ & $-0.675$ & $-5.356$ & $-6.369
$ & $0.0317$ \\ \hline
$K\eta \rightarrow \pi K$ & $-33.196$ & $78.965$ & $-0.460$ & $-15.275$ & $%
-1.373$ \\ 
$K\eta \rightarrow \pi K^{\ast }$ & $0.00425$ & $-18.683$ & $-5.715$ & $%
18.688$ & $-5.511$ \\ 
$K\eta \rightarrow \rho K$ & $6.819$ & $-5.863$ & $-0.871$ & $-0.955$ & $%
1.118$ \\ 
$K\eta \rightarrow \rho K^{\ast }$ & $0.288$ & $-0.413$ & $-15.160$ & $1.593$
& $-5.047$ \\ 
$K\eta \rightarrow \omega K$ & $3.124$ & $-2.091$ & $-0.936$ & $-1.032$ & $%
0.556$ \\ 
$K\eta \rightarrow \omega K^{\ast }$ & $-0.389$ & $-0.322$ & $-1.785$ & $%
1.182$ & $-0.625$ \\ \hline
$K^{\ast }\eta \rightarrow \pi K$ & $2.078$ & $0$ & $-$ & $-2.077$ & $-2.998$
\\ 
$K^{\ast }\eta \rightarrow \pi K^{\ast }$ & $0.078$ & $9.298$ & $-4.638$ & $%
0.514$ & $-36.521$ \\ 
$K^{\ast }\eta \rightarrow \rho K$ & $34.849$ & $10.952$ & $-15.036$ & $%
-16.308$ & $-2.838$ \\ 
$K^{\ast }\eta \rightarrow \rho K^{\ast }$ & $2.999$ & $-1.694$ & $-27.299$
& $1.181$ & $-5.701$ \\ 
$K^{\ast }\eta \rightarrow \omega K$ & $11.591$ & $3.936$ & $-15.140$ & $%
-5.276$ & $-2.807$ \\ 
$K^{\ast }\eta \rightarrow \omega K^{\ast }$ & $0.966$ & $-0.522$ & $-27.134$
& $0.409$ & $-5.647$ \\ \hline\hline
\end{tabular}%
\label{para}
\end{table}

The results shown in Figs. \ref{cross1} and \ref{cross2} can be conveniently
reproduced by parameterizing the squared matrix elements shown in Eq.(\ref%
{matrix}) according to 
\begin{equation}
|M_{n}|^{2}=a+b\left( \frac{\sqrt{s}}{\sqrt{s_{0}}}\right) ^{c}+d\left( 
\frac{\sqrt{s}}{\sqrt{s_{0}}}\right) ^{e}~{\rm mb}.  \label{parameter}
\end{equation}%
The dimensionless parameters $a$, $b$, $c$, $d$, and $e$ for the different
reactions considered in present study are given in Table \ref{para}.

\section{eta production in relativistic heavy ion collisions}

\label{heavyion}

In this section, we study the effect of eta absorption reactions studied in
previous sections on the time evolution of its abundance in heavy ion
collisions at RHIC. We describe the collision dynamics in a schematic
hydrodynamic model and assume that the hadronic matter is in thermal
equilibrium throughout the collision. Furthermore, we assume that pions, rho
mesons, omega mesons, and kaons as well as their resonances are in chemical
equilibrium during the collision because of their larger interaction cross
sections. For eta mesons, we consider two scenarios for their abundance in
the beginning of the hadronic stage after hadronization of the quark-gluon
plasma, i.e., they are either taken to be in chemical equilibrium with other
hadrons or completely absent. Their abundance during the evolution of the
subsequent hadronic matter is then described by a kinetic model that takes
into account their absorption and regeneration. We also compare the
final eta number in the two scenarios with that from assuming that eta
mesons remain in chemical equilibrium throughout the evolution of
the hadronic matter.

\subsection{rate equation}

The density $n_\eta$ of eta mesons changes in time according to the
following rate equation \cite{xia}: 
\begin{eqnarray}
\partial_\mu(n_\eta u^\mu)=\Psi,
\end{eqnarray}
where $u^\mu=\gamma(1,\mathbf{v})$ is the four velocity of the hadronic
matter fluid element of velocity $\mathbf{v}$ and $\gamma$ denotes the
Lorentz factor. The source term $\Psi$ is given by 
\begin{eqnarray}
\Psi =-\sum_{a,b,c}\langle\sigma_{a\eta\rightarrow bc}v\rangle n_\eta n_a
+\sum_{a,b,c}\langle\sigma_{bc\rightarrow a\eta}v\rangle n_b n_c ,
\label{rate}
\end{eqnarray}
where $n_a$, $n_b$, and $n_c$ are densities of meson types $a$, $b$, and $c$%
, respectively. The thermal averaged eta absorption and production cross
sections are denoted by $\langle\sigma_{a\eta\rightarrow bc}v\rangle$ and $%
\langle\sigma_{bc\rightarrow a\eta}v\rangle$, respectively, with $v$ the
relative velocity of initial two interacting particles. The thermal averaged
eta production cross sections are related to those of eta absorption cross
sections by 
\begin{eqnarray}
\langle\sigma_{a\eta\rightarrow bc}v\rangle n^{\mathrm{eq}}_a n^{\mathrm{eq}%
}_\eta= \langle\sigma_{bc\rightarrow a\eta}v\rangle n^{\mathrm{eq}}_b n^{%
\mathrm{eq}}_c,
\end{eqnarray}
where $n^{\mathrm{eq}}$ denotes the equilibrium density, i.e., 
\begin{equation}
n^{\mathrm{eq}}=\frac{dm^2T}{2\pi^2}K_2(m/T).
\end{equation}
In the above, $d$ and $m$ are the degeneracy and mass of a hadron, $T$ is
the temperature of the hadronic matter, and $K_2$ is the modified Bessel
function of the second kind. Using the above relation, the rate equation can
be written as 
\begin{eqnarray}
\partial_\mu(n_\eta u^\mu)=-\sum_{a,b,c}\langle \sigma_{a\eta\rightarrow
bc}v\rangle n^{\mathrm{eq}}_a(n_\eta-n^{\mathrm{eq}}_\eta).
\end{eqnarray}

\subsection{thermal averaged eta absorption cross sections}

With particle momenta in the hadronic matter approximated by the Boltzmann
distributions, the thermal averaged cross sections can be expressed as \cite%
{xia} 
\begin{eqnarray}
\langle\sigma v\rangle&=&[4\alpha^2_1 K_2(\alpha_1) \alpha^2_2
K_2(\alpha_2)]^{-1}  \nonumber \\
&&\times\int^\infty_{z_0} dz[z^2-(\alpha_1+\alpha_2)^2]
[z^2-(\alpha_1-\alpha_2)^2]K_1(z)\sigma(s=z^2 T^2),
\end{eqnarray}
with $\alpha_i=m_i/T$, $z_0=\mathrm{max}(\alpha_1+\alpha_2,\alpha_3
+\alpha_4)$, and $K_1$ being the modified Bessel function of the first kind.

\begin{figure}[ht]
\begin{minipage}{0.32 \textwidth}
\includegraphics[width=3.0in,height=2.5in,angle=270]{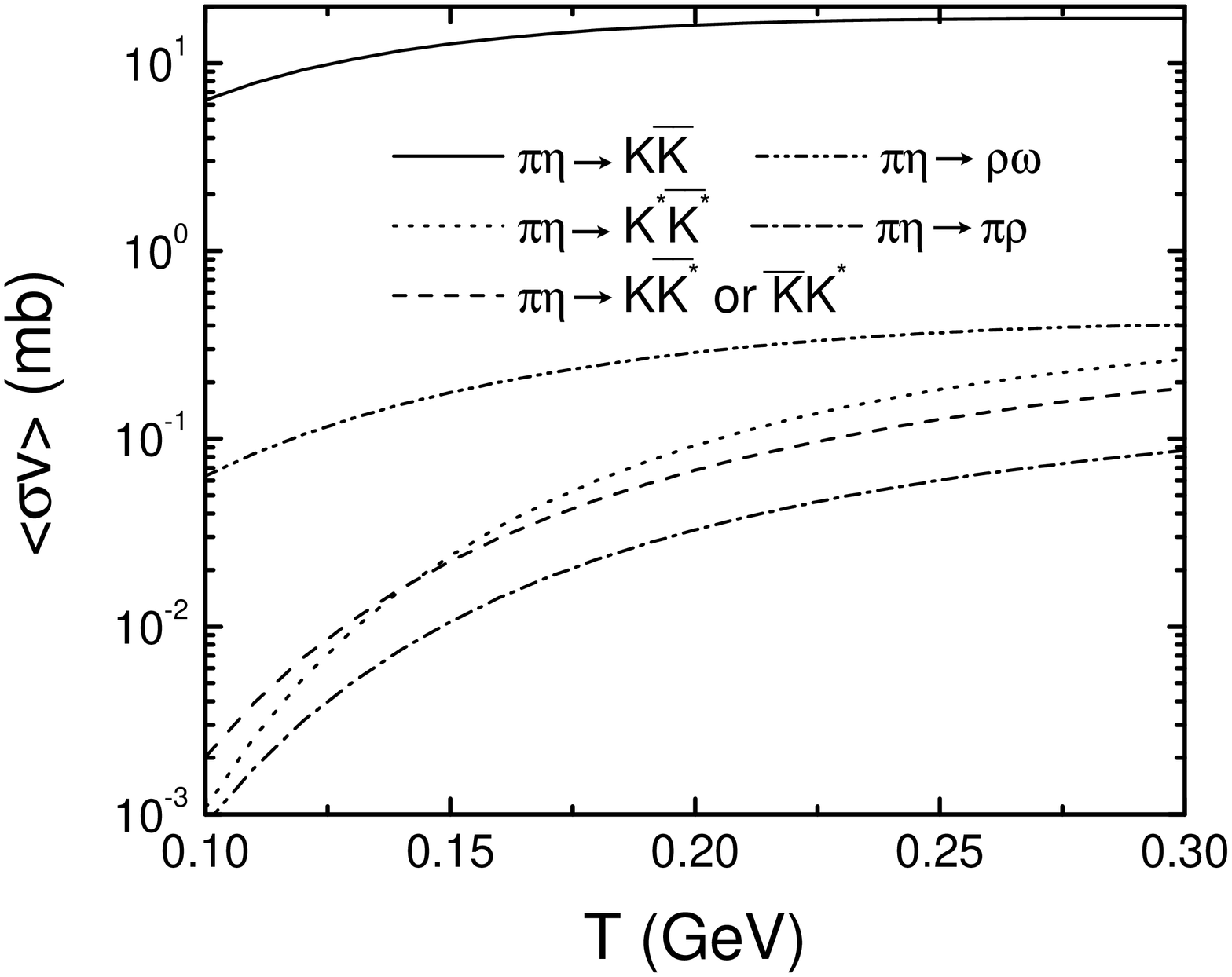}
\end{minipage}
\begin{minipage}{0.32 \textwidth}
\includegraphics[width=3.0in,height=2.5in,angle=270]{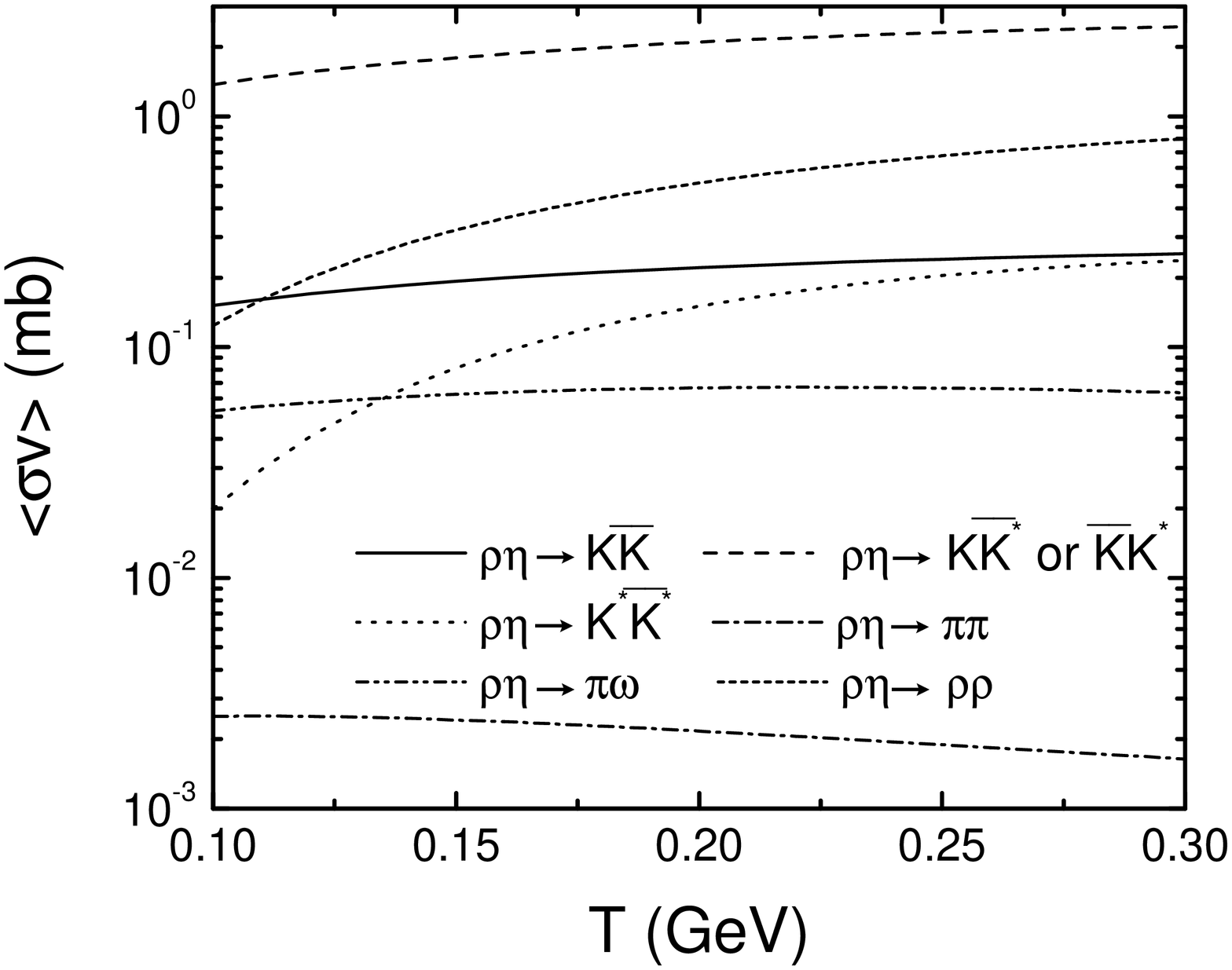}
\end{minipage}
\begin{minipage}{0.32 \textwidth}
\includegraphics[width=3.0in,height=2.5in,angle=270]{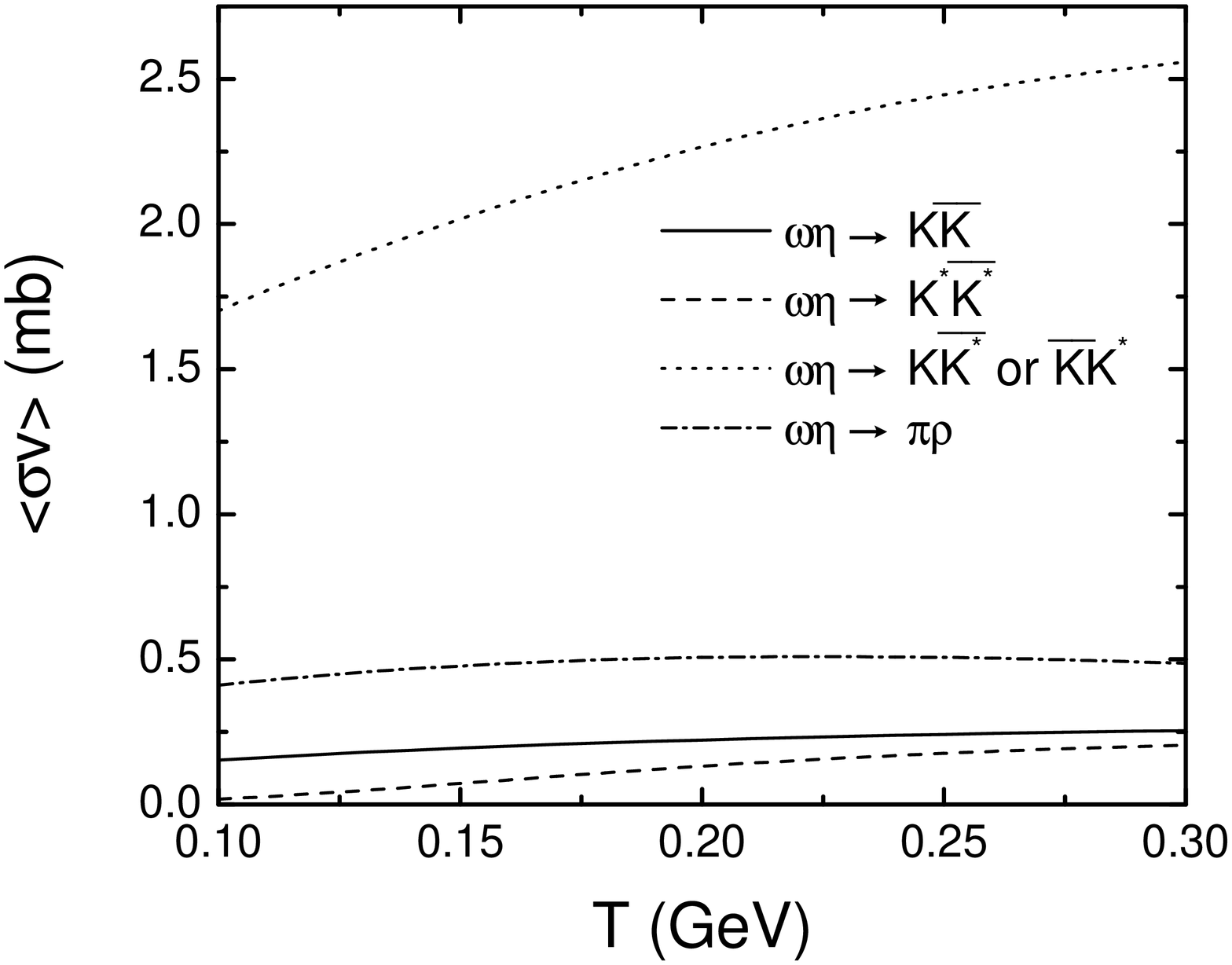}
\end{minipage}
\vspace{0.5cm}
\caption{Thermal averaged $\protect\eta$ absorption cross sections by pion
(left panel), rho (middle panel) and omega (right panel) mesons as functions
of temperature.}
\label{sigv1}
\end{figure}

\begin{figure}[ht]
\begin{minipage}{0.48 \textwidth}
\includegraphics[width=3.0in,height=3.0in,angle=270]{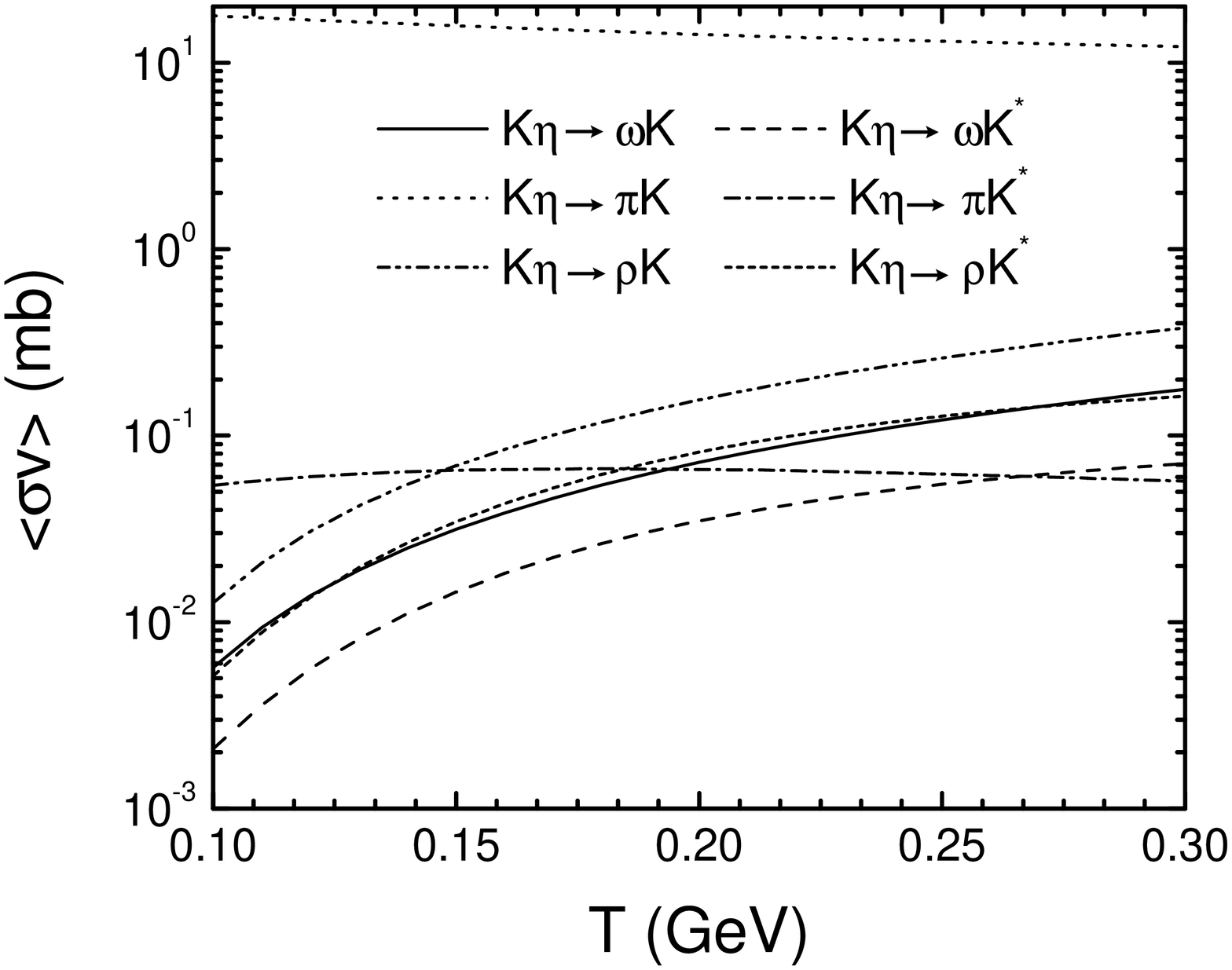}
\end{minipage}
\begin{minipage}{0.48 \textwidth}
\includegraphics[width=3.0in,height=3.0in,angle=270]{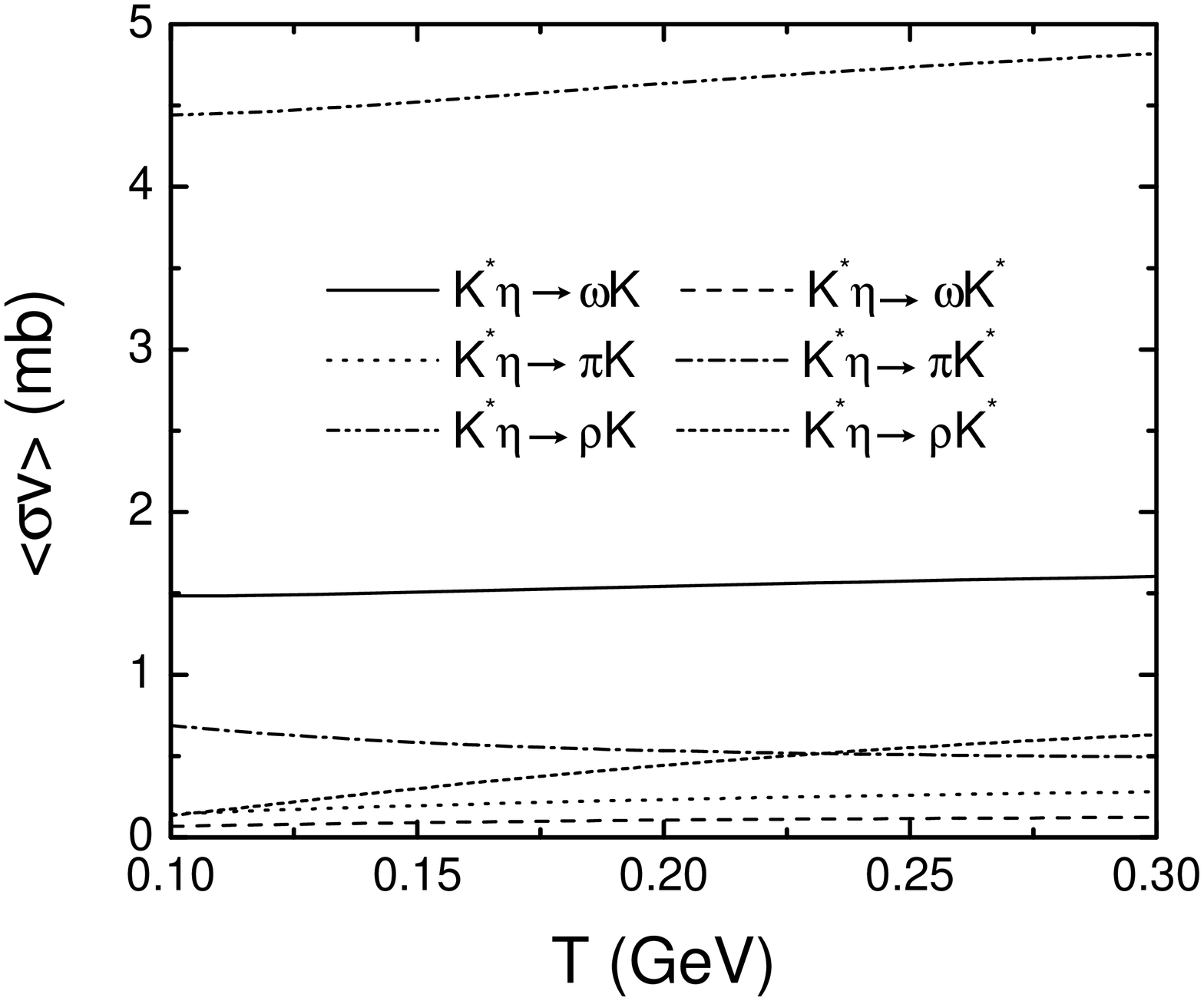}
\end{minipage}
\vspace{0.5cm}
\caption{Thermal averaged $\protect\eta$ absorption cross sections by $K$
(left panel) and $K^*$ (right panel) as functions of temperature.}
\label{sigv3}
\end{figure}

The resulting temperature dependence of $\langle\sigma v\rangle$ is shown in
Fig. \ref{sigv1} for the reactions of $\pi\eta$, $\rho\eta$, and $\omega\eta$
and in Fig. \ref{sigv3} for the reactions of $K\eta$ and $K^*\eta$. It is
seen that the thermal averaged cross sections for most reactions increase
with increasing temperature. As for cross sections, the reactions $%
\pi\eta\to K\bar K$ and $K\eta\to\pi K$ have largest thermal 
averaged cross sections.

\subsection{collision dynamics at RHIC}

Since the particle distribution in central heavy ion collisions at RHIC is
approximately uniform in midrapidity and the geometry of the collision is
cylindrically symmetric, it is convenient to use the cylindrical coordinates 
$r$, $\varphi$, $\tau$, and $\eta$ with the latter given by 
\begin{eqnarray}
\tau=\sqrt{t^2-z^2},\ \ \eta=\frac{1}{2}\ln\frac{t+z}{t-z}.
\end{eqnarray}
Assuming longitudinal boost invariance and allowing for radial transverse
expansion, then one has $u^{\eta}=u^{\varphi}=0$. For uniform density
distribution in the transverse plane, averaging over the radial coordinate
gives 
\begin{eqnarray}  \label{kinetic}
\frac{1}{\tau R^2(\tau)}\frac{\partial}{\partial\tau}(\tau R^2(\tau)n_\eta
\langle u^\tau\rangle)= -\sum_{a,b,c}\langle\sigma_{\eta a\rightarrow bc}v
\rangle n^{eq}_a(n_\eta-n^{eq}_\eta)
\end{eqnarray}
In the above, $R(\tau)$ is the transverse radius of the system and $\langle
u^\tau\rangle$ is the averaged $\tau$ component of the four velocity and is
given by 
\begin{eqnarray}
\langle u^\tau \rangle = \frac{2}{R^2 (\tau)} \int_0^{R(\tau)} dr\, r
u^\tau(r)\,.
\end{eqnarray}

At midrapidity, the four velocity of hadronic fluid element $u^\mu$ can be
expressed in terms of the radial flow velocity $\beta_r$ as 
\begin{eqnarray}
u^\tau = \gamma_r = \frac{1}{\sqrt{1 -\beta_r^2}} \,.
\end{eqnarray}
With the usual ansatz for the radial velocity, i.e., 
\begin{eqnarray}
\beta_r (\tau,r) = \frac{dR}{d\tau}\left( \frac{r}{R} \right)\,,
\end{eqnarray}
we have 
\begin{eqnarray}
\langle u^\tau \rangle = \int_0^1 dy\,\frac{1}{\sqrt{1-(dR/d\tau)^2 y}}.
\end{eqnarray}

To determine the time evolution of the transverse radius of the fireball, we
follow the model used in Ref.\cite{chen}. Since we are interested in the
time evolution of the eta abundance during the hadronic phase of the
collision, we start from the end of the mixed phase $\tau_H$ and write 
\begin{equation}
R(\tau)=R_H+v_H(\tau-\tau_H)+\frac{a}{2}(\tau-\tau_H)^2.
\end{equation}
In the above, $R_H\approx 9$ fm and $v_H\approx 0.4c$ are, respectively, the
transverse radius and flow velocity of the fireball at $\tau_H=7.5$ fm/$c$,
while $a=0.02c^2$/fm is the acceleration in the transverse expansion. Values
of these parameters are determined from fitting the measured transverse
energy $\simeq 788$ GeV as well as the extracted freeze out temperature $%
T_F=125$ MeV and transverse flow velocity $\simeq 0.65c$ of midrapidity
hadrons in central Au+Au collisions at $\sqrt{s_{NN}}=200$ GeV. Assuming
that the hadronic matter expands isentropically, the time dependence of the
temperature of the fireball obtained in Ref.\cite{chen} can be parameterized
as 
\begin{eqnarray}
T(\tau)=T_C-(T_H-T_F)\left(\frac{\tau-\tau_H}{\tau_F-\tau_H}\right)^{0.8},
\end{eqnarray}
where $T_H$ is the temperature of the hadronic matter at the end of the
mixed phase and is thus the same as the critical temperature $T_C$ for the
quark-gluon plasma to hadronic matter transition. As in Ref.\cite{chen}, we
take $T_H=T_C=175$ MeV. The freeze out temperature $T_F=125$ MeV then leads
to a freeze out time $\tau_F\approx 17.3$ fm/$c$.

\subsection{time evolution of the eta abundance}

\begin{figure}[ht]
\begin{minipage}{0.32 \textwidth}
\includegraphics[width=3.0in,height=2.5in,angle=270]{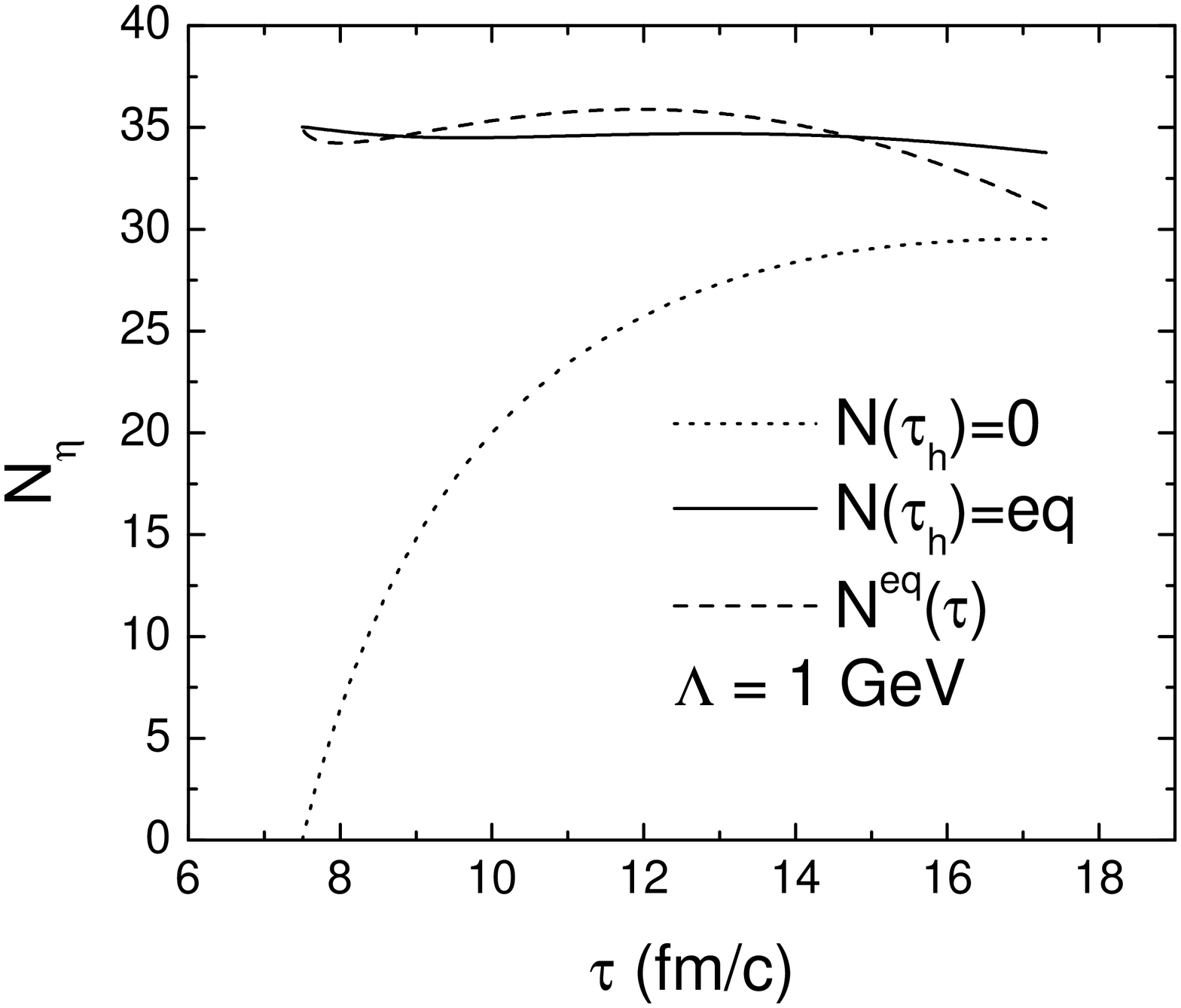}
\end{minipage}
\begin{minipage}{0.32 \textwidth}
\includegraphics[width=3.0in,height=2.5in,angle=270]{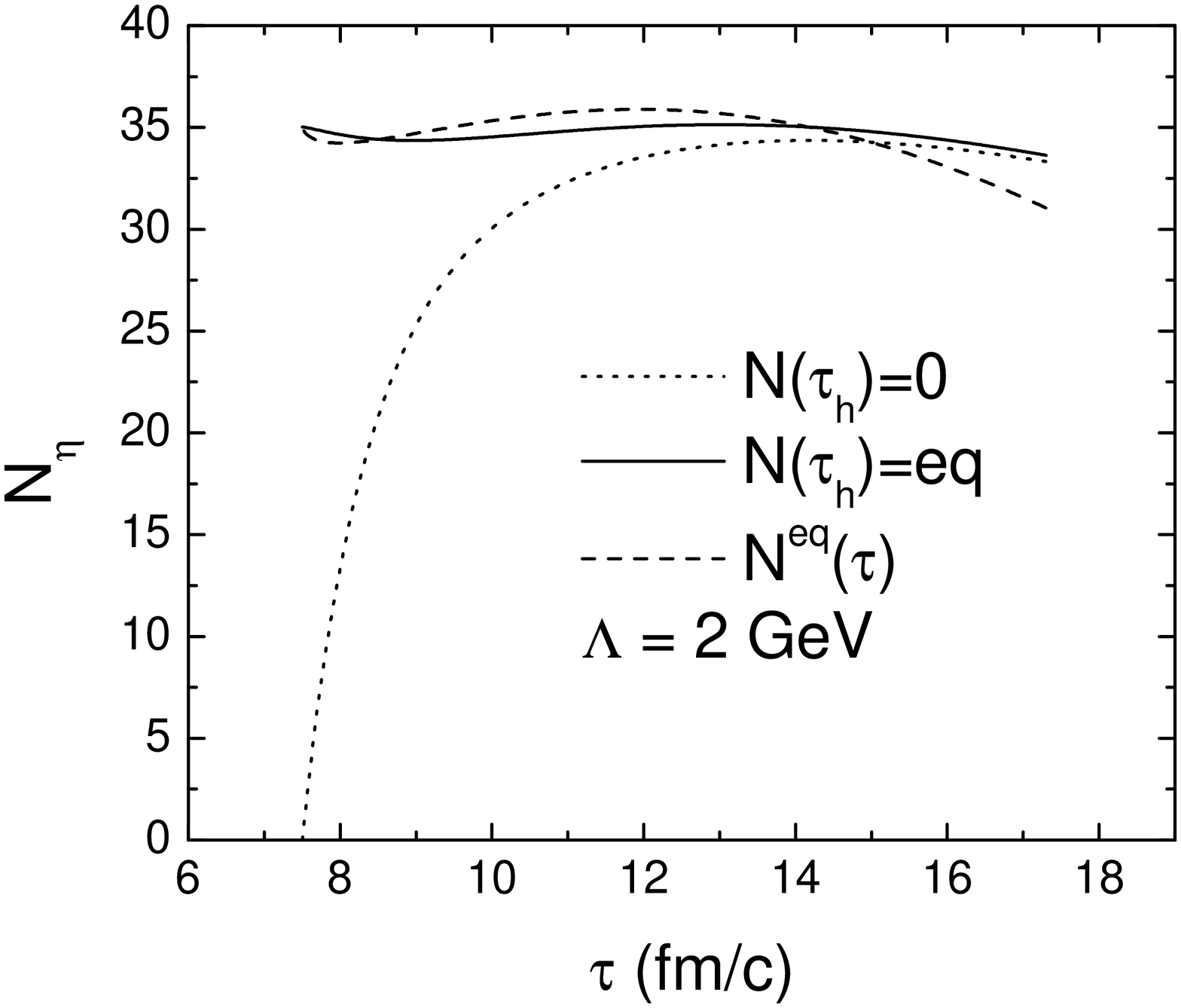}
\end{minipage}
\begin{minipage}{0.32 \textwidth}
\includegraphics[width=3.0in,height=2.5in,angle=270]{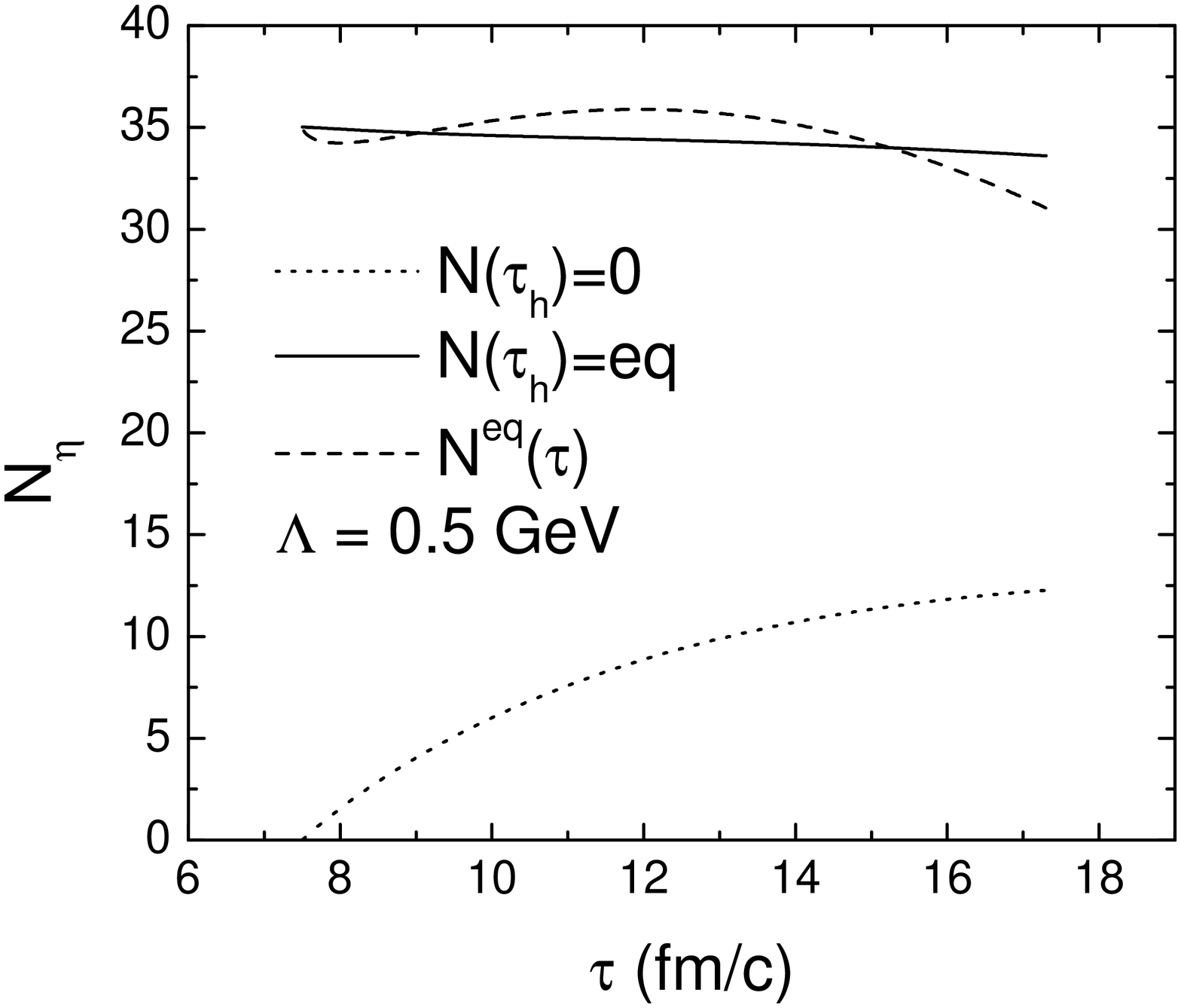}
\end{minipage}
\vspace{0.5cm}
\caption{Time dependence of the abundance of midrapidity $\protect\eta$
mesons in the hot hadronic gas formed from central Au+Au collisions at $%
\protect\sqrt{s_{NN}}=200$ GeV at RHIC for cutoff parameter $\Lambda=1$ GeV
(left panel), 2 GeV (middle panel), and 0.5 GeV (right panel). Solid and
dotted lines correspond, respectively, to eta mesons that are chemically 
equilibrated or absent at the beginning of the hadronic phase, while 
dashed lines correspond to eta mesons that are always in chemical 
equilibrium.}
\label{num}
\end{figure}

To study how the eta meson abundance evolves in time, we consider two
scenarios for the initial eta meson number, i.e., no eta meson is present at 
$\tau_H$ or the eta meson is in chemical equilibrium with other hadrons at $%
\tau_H$ as in the statistical model \cite{braun}. In the first case,
subsequent hadronic interactions increase the final eta meson number to
about 30 at freeze out, as shown by the dotted line in the left panel of
Fig. \ref{num}. This scenario is, however, unrealistic as we expect eta
mesons to be appreciably produced during hadronization of the quark-gluon
plasma. For example, in the quark coalescence model \cite{greco}, the number
of eta mesons produced at hadronization ranges from 12 for a small eta meson
root mean square radius of 0.5 fm to 37 for a larger radius of 1 fm. In the
second scenario of equilibrated eta mesons, the initial eta meson number is $%
N(\tau_H)\approx 35$. Including subsequent hadronic absorption and
regeneration does not change much the eta abundance, and its number at
freeze out is decreased slightly to about 34 as shown by the solid line in
the left panel of Fig. \ref{num}. This result is consistent with the
assumption of the statistical model that hadron abundances in heavy ion
collisions at RHIC are largely determined at hadronization. For comparison,
we have also shown in the left panel of Fig. \ref{num} by the dashed line
results from the assumption that eta mesons are always in chemical
equilibrium with other hadrons during evolution of the hadronic matter. 
In this case, the eta number shows a small initial increase that is followed 
by a decrease to about 32 at freeze out. The similarity among the final eta 
abundance in all three scenarios indicates that the yield of eta meson in 
relativistic heavy ion collisions is not very sensitive to its production 
mechanism. This is not surprising as the cross sections for the
reactions $\pi\eta\to K\bar K$ and $K\eta\to\pi K$ obtained with a cutoff 
parameter $\Lambda=1$ GeV in the form factor are sufficient large, implying 
that eta mesons are strongly interacting and thus likely reach both 
thermal and chemical equilibrium in the hot hadronic matter produced 
at relativistic heavy ion collisions. For a larger cutoff parameter of 
$\Lambda=2$ GeV in the form factor, these cross sections will be even 
larger, and we would expect an even more similar final eta abundance
for the three scenarios, and this is indeed seen in the middle panel 
of Fig. \ref{num}. Even for a smaller cutoff parameter of $\Lambda=0.5$ 
GeV, which leads to a much smaller cross sections for eta absorption and 
production, final state interactions in the hadronic matter can still 
keep eta mesons close to chemical equilibrium if they are initially in 
chemical equilibrium as shown in the right panel of Fig. \ref{num}. 
On the other hand, the final abundance of eta mesons will be 
significantly below the equilibrium value if they are initially
absent in the hadronic matter, which is, however, unlikely as we 
have commented in the above.

\section{summary}

\label{summary}

Knowledge of eta meson interactions in hadronic matter is not only of
interest in its own right but also important for extracting information on
the properties of the hot dense matter formed in high energy heavy ion
collisions. Since there is no empirical information on the absorption cross
sections of the eta meson with pion, rho and omega mesons as well as kaon
and its resonance, which are the most abundant particles in high energy
heavy ion collisions, we have evaluated these cross sections based on the
tree-level diagrams from the $[SU(3)_{\mathrm{L}} \times SU(3)_{\mathrm{R}%
}]_{\mathrm{global}} \times [SU(3)_V]_{\mathrm{local}}$ chiral Lagrangian
with hidden local symmetry and including symmetry breaking effects. Using
empirical hadron masses and coupling constants as well as reasonable
values for the cutoff parameters in the form factors at interaction
vertices, we find that although cross sections for most eta absorption
reactions by mesons are less than 1 mb, the reactions $\pi\eta\to K\bar K$
and $K\eta\to\pi K$ are more than 10 mb. To see the effects of these 
reactions on the yield of eta mesons in relativistic heavy ion 
collisions, we have solved a kinetic equation, based on a schematic
model for the dynamics of heavy ion collisions, to follow 
the time evolution of the abundance of eta mesons. We find that the
final abundance of eta mesons at freeze out of the hadronic matter
is close to chemical equilibrium, irrespective of the initial value 
at hadronization of the quark-gluon plasma. In particular, if eta 
mesons are initially in chemical equilibrium with other hadrons,  
their number is not strongly affected by their subsequent
interactions during the expansion of the hadronic matter. 

\begin{acknowledgments}
This work was supported in part by the US National Science Foundation under
Grant Nos. PHY-0098805 and PHY-0457265 and the Welch Foundation under Grant
No. A-1358 (C.M.K. and W.L.) as well as by the National Science Foundation
of China under Grant Nos. 10105008 and 10575071 (L.W.C.).
\end{acknowledgments}

\appendix

\section{eta absorption by pion}

\label{pion}

Using subscripts 1 and 2 to denote the initial-state particles and
3 and 4 for the final-state particles in the order from left to right 
in all the Feynman diagrams shown in Figs. \ref{dpion}-\ref{dkstar}, 
$\epsilon_{i\mu}$ for the polarization vector of vector mesons, and 
also the usual Mandelstam variables $s=(p_1+p_2)^2$, $t=(p_1 -p_3)^2$ 
and $u=(p_1 -p_4)^2$, the amplitudes for the absorption by mesons
can be written explicitly as given below.  For propagators, we do not
include the width of exchanged particles as its effect is negligible 
due to the large threshold of the reactions. We note these expressions 
are obtained with $a=2$ in the chiral Lagrangian.   

\subsection{$\protect\pi\protect\eta\to K\bar K$}

The amplitude for this reaction is given by 
\begin{eqnarray}
\mathcal{M}_{\pi\eta\to K\bar K}=\tau^a_{ij}(\mathcal{M}_{1a}+\mathcal{M}%
_{1b} +\mathcal{M}_{1c}),
\end{eqnarray}
with 
\begin{eqnarray}
\mathcal{M}_{1a} &=& \frac{g^2}{\sqrt{6}}\frac{(1+c_V)^2}{(1+c_A) 
\sqrt{1+\frac{2}{3}c_A}}(p_1 +p_3)_\mu\frac{1}{t-m_{K^*}^2}  \nonumber \\
&&\times\left[-g^{\mu\nu} +\frac{(p_1 -p_3)^\mu (p_1 -p_3)^\nu}{m_{K^*}^2}%
\right](p_2 +p_4 )_\nu,  \nonumber \\
\mathcal{M}_{1b} &=& \frac{g^2}{\sqrt{6}}\frac{(1+c_V)^2}{(1+c_A) 
\sqrt{1+\frac{2}{3}c_A}}(p_1 +p_4)_\mu\frac{1}{u-m_{K^*}^2}  \nonumber \\
&&\times\left[-g^{\mu\nu}+\frac{(p_1 -p_4 )^\mu (p_1 -p_4)^\nu}{m_{K^*}^2}%
\right](p_2 +p_3 )_\nu,  \nonumber \\
\mathcal{M}_{1c} &=&\frac{1}{3\sqrt{6}f_\pi^2}\frac{1}{(1+c_A) 
\sqrt{1+\frac{2}{3}c_A}}\left[\left(1+\frac{3}{2}c_A \right)
(p_1\cdot p_4+p_1\cdot p_3)\right. \nonumber \\
&&\left.+p_2\cdot p_3+p_2\cdot p_4+\left(2+3c_A\right)p_1\cdot p_2 
+2p_3\cdot p_4+m_\pi^2\right].
\end{eqnarray}

\subsection{$\protect\pi\protect\eta\to K\bar K^*(\bar K K^*)$}

The amplitude for this reaction is given by 
\begin{eqnarray}
\mathcal{M}_{\pi\eta\to K\bar K^*(\bar KK^*)}=\tau^a_{ij}(\mathcal{M}%
_{2a}^\mu +\mathcal{M}_{2b}^\mu)\epsilon_{4\mu},
\end{eqnarray}
with 
\begin{eqnarray}
\mathcal{M}_{2a}^\mu&=&\frac{gg_{K^*K^*\eta}}{\sqrt{6}}
\frac{(1+c_V)c_{\rm wz}}{\sqrt{(1+c_A)(1+\frac{2}{3}c_A)}}
(p_1+p_3)^\nu\nonumber\\
&&\times\left[-g_{\nu\nu^{\prime}}+\frac{(p_1-p_3)_\nu
(p_1-p_3)_{\nu^{\prime}}}{m^2_{K^*}}\right]\frac{1}{t-m^2_{K^*}}
\epsilon^{\alpha\nu^{\prime}\beta\mu}p_{4\beta}(p_2-p_4)_\alpha, \nonumber\\
\mathcal{M}_{2b}^\mu&=&\frac{gg_{K^*K^*\pi}}{\sqrt{6}}\frac{(1+c_V)(1+2c_{%
\mathrm{wz}})} {\sqrt{(1+c_A)(1+\frac{2}{3}c_A)}}  \nonumber \\
&&\times\epsilon^{\alpha\nu\beta\mu}p_{4\beta}(p_1-p_4)_\alpha
\left[-g_{\nu\nu^{\prime}}+\frac{(p_1-p_4)_\nu(p_1-p_4)_{\nu^{\prime}}}{%
m^2_{K^*}} \right]\frac{1}{u-m^2_{K^*}}(p_2+p_3)^{\nu^{\prime}}.
\end{eqnarray}

\subsection{$\protect\pi\protect\eta\to K^*\bar K^*$}

The amplitude for this reaction is given by 
\begin{eqnarray}
\mathcal{M}_{\pi\eta\to K^*\bar K^*}=\tau^a_{ij}(\mathcal{M}_{3a}^{\mu\nu}+ 
\mathcal{M}_{3b}^{\mu\nu}+\mathcal{M}_{3c}^{\mu\nu}+\mathcal{M}%
_{3d}^{\mu\nu}) \epsilon_{3\mu}\epsilon_{4\nu},
\end{eqnarray}
with 
\begin{eqnarray}
\mathcal{M}_{3a}^{\mu\nu} &=& \frac{g^2}{\sqrt{6}}
\frac{(1+c_V)^2}{(1+c_A)\sqrt{1+\frac{2}{3}c_A}}(2p_1 -p_3)^\mu 
\frac{1}{t-m_K^2}(2p_2-p_4)^\nu,  \nonumber \\
\mathcal{M}_{3b}^{\mu\nu} &=& \frac{g^2}{\sqrt{6}}
\frac{(1+c_V)^2}{(1+c_A)\sqrt{1+\frac{2}{3}c_A}}(2p_1 -p_4)^\nu 
\frac{1}{u-m_K^2}(2p_2-p_3)^\mu,  \nonumber \\
\mathcal{M}_{3c}^{\mu\nu}&=&\frac{g_{K^*K^*\pi}g_{K^*K^*\eta}}{\sqrt{6}}
\frac{(1+2c_{\rm wz})c_{\rm wz}}{\sqrt{1+\frac{2}{3}c_A}}
\epsilon^{\alpha\mu\beta\gamma}p_{3\alpha}(p_3-p_1)_\beta\nonumber\\
&&\times\left[-g_{\gamma\gamma^{\prime}}+ \frac{%
(p_1-p_3)_\gamma (p_1-p_3)_{\gamma^{\prime}}}{m^2_{K^*}}\right]\frac{1}{%
t-m^2_{K^*}} \epsilon^{\alpha^{\prime}\gamma^{\prime}\beta^{\prime}\nu}p_{4%
\beta^{\prime}} (p_2-p_4)_{\alpha^{\prime}},  \nonumber \\
\mathcal{M}_{3d}^{\mu\nu}&=&\frac{g_{K^*K^*\pi}g_{K^*K^*\eta}}{\sqrt{6}}
\frac{(1+2c_{\rm wz})c_{\rm wz}}{\sqrt{1+\frac{2}{3}c_A}} 
\epsilon^{\alpha\gamma\beta\nu}p_{4\beta} (p_4-p_1)_\alpha\nonumber\\
&&\times\left[-g_{\gamma\gamma^{\prime}}+ \frac{%
(p_1-p_4)_\gamma(p_1-p_4)_{\gamma^{\prime}}}{m^2_{K^*}}\right] \frac{1}{%
u-m^2_{K^*}}\epsilon^{\alpha^{\prime}\mu\beta^{\prime}\gamma^{\prime}}
p_{3\alpha^{\prime}}(p_2-p_3)_{\beta^{\prime}}.
\end{eqnarray}

\subsection{$\protect\pi\protect\eta\to\protect\rho\protect\omega$}

The amplitude for this reaction is given by 
\begin{eqnarray}
\mathcal{M}_{\pi\eta\to\rho\omega}=\delta_{ab}(\mathcal{M}_{4a}^{\mu\nu} +%
\mathcal{M}_{4b}^{\mu\nu})\epsilon_{3\mu}\epsilon_{4\nu},
\end{eqnarray}
with 
\begin{eqnarray}
\mathcal{M}_{4a}^{\mu\nu}&=&\frac{g_{\rho\omega\pi}g_{\omega\omega\eta}}
{\sqrt{6}}\frac{1}{\sqrt{1+\frac{2}{3}c_A}} \epsilon^{\alpha\mu\beta\gamma}
p_{3\alpha}(p_3-p_1)_\beta  \nonumber \\
&\times&\left[-g_{\gamma\gamma^{\prime}}+\frac{(p_1-p_3)_\gamma
(p_1-p_3)_{\gamma^{\prime}}}{m^2_\rho}\right]\frac{1}{t-m^2_\rho}
\epsilon^{\alpha^{\prime}\gamma^{\prime}\beta^{\prime}\nu}p_{4\beta^{%
\prime}}(p_2-p_4)_{\alpha^{\prime}},  \nonumber \\
\mathcal{M}_{4b}^{\mu\nu}&=&\frac{g_{\rho\omega\pi}g_{\rho\rho\eta}}{\sqrt{6}}
\frac{1}{\sqrt{1+\frac{2}{3}c_A}} \epsilon^{\alpha\gamma\beta\nu}
p_{4\beta}(p_4-p_1)_\alpha  \nonumber \\
&\times&\left[-g_{\gamma\gamma^{\prime}}+\frac{(p_1-p_4)_\gamma
(p_1-p_4)_{\gamma^{\prime}}}{m^2_\rho}\right]\frac{1}{u-m^2_\rho}
\epsilon^{\alpha^{\prime}\mu\beta^{\prime}\gamma^{\prime}}p_{3\alpha^{%
\prime}}(p_2-p_3)_{\beta^{\prime}}.
\end{eqnarray}

\subsection{$\protect\pi\protect\eta\to\protect\pi\protect\rho$}

The amplitude for this reaction is given by 
\begin{eqnarray}
\mathcal{M}_{\pi\eta\to\pi\rho}=i\epsilon_{abc}\mathcal{M}%
_5^\mu\epsilon_{4\mu},
\end{eqnarray}
with 
\begin{eqnarray}
\mathcal{M}_5^\mu&=&\frac{gg_{\rho\rho\eta}}{\sqrt{6}}
\frac{1}{\sqrt{1+\frac{2}{3}c_A}}(p_1+p_3)^\nu\nonumber\\
&&\times\left[-g_{\nu\nu^{\prime}}+\frac{(p_1-p_3)_\nu
(p_1-p_3)_{\nu^{\prime}}}{m^2_\rho}\right]\frac{1}{t-m^2_\rho} 
\epsilon^{\alpha\nu^{\prime}\beta\mu}p_{4\beta}(p_2-p_4)_\alpha.
\end{eqnarray}

\section{eta absorption by rho meson}

\label{rho}

\subsection{$\protect\rho\protect\eta\to K\bar K$}

The amplitude for this reaction is given by 
\begin{eqnarray}
\mathcal{M}_{\rho\eta\to K\bar K}=\tau^a_{ij}(\mathcal{M}_{6a}^\mu +\mathcal{%
M}_{6b}^\mu+\mathcal{M}_{6c}^\mu)\epsilon_{1\mu},
\end{eqnarray}
with 
\begin{eqnarray}
\mathcal{M}_{6a}^\mu&=&\frac{gg_{\rho K^*K}}{\sqrt{6}}
\frac{1+c_V}{(1+c_A) \sqrt{1+\frac{2}{3}c_A}}\epsilon^{\nu\mu\alpha\beta}
p_{1\nu}(p_3-p_1)_\alpha  \nonumber \\
&&\times\left[-g_{\beta\beta^{\prime}}+\frac{(p_1-p_3)_\beta(p_1-p_3)_{\beta^{%
\prime}}} {m^2_{K^*}}\right]\frac{1}{t-m^3_{K^*}}(p_2+p_4)^{\beta^{\prime}}, 
\nonumber \\
\mathcal{M}_{6b}^\mu&=&\frac{gg_{\rho K^*K}}{\sqrt{6}}
\frac{1+c_V}{(1+c_A) \sqrt{1+\frac{2}{3}c_A}}\epsilon^{\nu\mu\alpha\beta}
p_{1\nu}(p_1-p_4)_\alpha  \nonumber \\
&&\times\left[-g_{\beta\beta^{\prime}}+\frac{(p_1-p_4)_\beta
(p_1-p_4)_{\beta^{\prime}}}{m^2_{K^*}}\right]\frac{1}{u-m^2_{K^*}}
(p_2+p_3)^{\beta^{\prime}},  \nonumber \\
\mathcal{M}_{6c}^\mu&=&\frac{gg_{\rho\rho\eta}}{2\sqrt{6}}\frac{1}{(1+c_A)
\sqrt{1+\frac{2}{3}c_A}}
\epsilon^{\nu\mu\alpha\beta}p_{1\nu}(p_1+p_2)_\alpha  \nonumber \\
&&\times\left[-g_{\beta\beta^{\prime}}+\frac{(p_1+p_2)_\beta(p_1+p_2)_{\beta^{%
\prime}}} {m^2_\rho}\right]\frac{1}{s-m^2_\rho}(p_4-p_3)^{\beta^{\prime}}.
\end{eqnarray}

\subsection{$\protect\rho\protect\eta\to K\bar K^*(\bar KK^*)$}

The amplitude for this reaction is given by 
\begin{eqnarray}
\mathcal{M}_{\rho\eta\to K\bar K^*(\bar KK^*)}=\tau^a_{ij}(\mathcal{M}%
_{7a}^{\mu\nu} +\mathcal{M}_{7b}^{\mu\nu}+\mathcal{M}_{7c}^{\mu\nu} +%
\mathcal{M}_{7d}^{\mu\nu}+\mathcal{M}_{7e}^{\mu\nu})
\epsilon_{1\mu}\epsilon_{4\nu},
\end{eqnarray}
with 
\begin{eqnarray}
\mathcal{M}_{7a}^{\mu\nu} &=& \frac{g^2 (1+c_V)} {\sqrt{6}%
\sqrt{(1+c_A)^3}\sqrt{1+\frac{2}{3}c_A}} (2p_3-p_1)^\mu\frac{1}{t-m_K^2}%
(2p_2-p_4)^\nu,  \nonumber \\
\mathcal{M}_{7b}^{\mu\nu} &=& \frac{g^2 (1+c_V)}{\sqrt{6}\sqrt{%
(1+c_A)(1+\frac{2}{3}c_A)}}[(2p_4-p_1)^\mu g^{\nu\alpha} +(2p_1-p_4)^\nu
g^{\mu\alpha}-(p_4 +p_1)^{\alpha}g^{\mu\nu}]  \nonumber \\
&&\times\frac{1}{u-m^2_{K^*}}\left[-g_{\alpha\beta} +\frac{(p_1
-p_4)_{\alpha}(p_1 -p_4)_{\beta}} {m_{K^*}^2}\right](p_2 +p_3)^{\beta}, 
\nonumber \\
\mathcal{M}_{7c}^{\mu\nu}&=&\frac{g^2_{K^*K^*\eta}}{\sqrt{6}}
\frac{c_{\rm wz}}{\sqrt{(1+c_A) (1+\frac{2}{3}c_A)}}
\epsilon^{\alpha\mu\beta\gamma}p_{1\alpha}(p_3-p_1)_\beta\nonumber\\
&&\times\left[-g_{\gamma\gamma^{\prime}}+\frac{(p_1-p_3)_\gamma
(p_1-p_3)_{\gamma^{\prime}}%
} {m^2_{K^*}}\right]\frac{1}{t-m^2_{K^*}}\epsilon^{\alpha^{\prime}\nu\beta^{%
\prime}\gamma^{\prime}} p_{4\alpha^{\prime}}(p_4-p_2)_{\beta^{\prime}}, 
\nonumber \\
\mathcal{M}_{7d}^{\mu\nu}&=&-\frac{g_{\rho\rho\eta}g_{\rho K^*K}}
{2\sqrt{6}}\frac{1}{\sqrt{(1+c_A) (1+\frac{2}{3}c_A)}}
\epsilon^{\alpha\mu\beta\gamma}p_{1\alpha}(p_1+p_2)_\beta\nonumber \\
&&\times\left[-g_{\gamma\gamma^{\prime}}+\frac{(p_1+p_2)_\gamma(p_1+p_2)_{%
\gamma^{\prime}}} {m^2_\rho}\right]\frac{1}{s-m^2_\rho} 
\epsilon^{\alpha^{\prime}\nu\beta^{\prime}\gamma^{\prime}}
p_{4\alpha^{\prime}}(p_3+p_4)_{\beta^{\prime}},  \nonumber \\
\mathcal{M}_{7e}^{\mu\nu} &=& \frac{g^2}{\sqrt{6}}\frac{c_V} {\sqrt{(1+c_A
)(1+\frac{2}{3}c_A)}} g^{\mu\nu}.
\end{eqnarray}

\subsection{$\protect\rho\protect\eta\to K^*\bar K^*$}

The amplitude for this reaction is given by 
\begin{eqnarray}
\mathcal{M}_{\rho\eta\to K^*\bar K^*}=\tau^a_{ij}(\mathcal{M}%
_{8a}^{\mu\nu\alpha} +\mathcal{M}_{8b}^{\mu\nu\alpha}+\mathcal{M}%
_{8c}^{\mu\nu\alpha} +\mathcal{M}_{8d}^{\mu\nu\alpha}+\mathcal{M}%
_{8e}^{\mu\nu\alpha}) \epsilon_{1\mu}\epsilon_{3\nu}\epsilon_{4\alpha},
\end{eqnarray}
with 
\begin{eqnarray}
\mathcal{M}_{8a}^{\mu\nu\alpha}&=&\frac{gg_{\rho K^*K}}{\sqrt{6}} 
\frac{1+c_V}{(1+c_A)\sqrt{1+\frac{2}{3}c_A}}
\epsilon^{\beta\mu\gamma\nu}p_{1\beta} p_{3\gamma}\frac{1}{t-m^2_K}
(2p_2-p_4)^{\alpha},  \nonumber \\
\mathcal{M}_{8b}^{\mu\nu\alpha}&=&\frac{gg_{\rho K^*K}}{\sqrt{6}} 
\frac{1+c_V}{(1+c_A)\sqrt{1+\frac{2}{3}c_A}}
\epsilon^{\beta\mu\gamma\alpha}p_{1\beta} p_{4\gamma}\frac{1}{u-m^2_K}
(p_3-2p_2)^{\nu},  \nonumber \\
\mathcal{M}_{8c}^{\mu\nu\alpha}&=& \frac{gg_{K^*K^*\eta}}{\sqrt{6}}
\frac{c_{\rm wz}} {\sqrt{(1+\frac{2}{3}c_A)}}  \nonumber \\
&&\times[(2p_1-p_3)^\nu g^{\mu\gamma}+(2p_3 -p_1)^\mu g^{\gamma\nu}
-(p_1+p_3)^\gamma g^{\mu\nu}]  \nonumber \\
&&\times\frac{1}{t-m^2_{K^*}}\left[-g_{\gamma\gamma^{\prime}} +\frac{(p_1
-p_3)_\gamma(p_1 -p_3)_{\gamma^{\prime}}}{m_{K^*}^2}\right]
\epsilon^{\beta\gamma^{\prime}\lambda\alpha}p_{4\lambda}(p_2 -p_4)_{\beta}, 
\nonumber \\
\mathcal{M}_{8d}^{\mu\nu\alpha}&=&\frac{gg_{K^*K^*\eta}}{\sqrt{6}}
\frac{c_{\rm wz}} {\sqrt{(1+\frac{2}{3}c_A)}}  \nonumber \\
&&\times[(2p_1-p_4)^\alpha g^{\mu\gamma}+(2p_4 -p_1)^\mu
g^{\gamma\alpha}-(p_1+p_4)^\gamma g^{\mu\alpha}]  \nonumber \\
&&\times\frac{1}{u-m^2_{K^*}}\left[-g_{\gamma\gamma^{\prime}} +\frac{(p_1
-p_4)_\gamma(p_1 -p_4)_{\gamma^{\prime}}}{m_{K^*}^2}\right]
\epsilon^{\beta\gamma^{\prime}\lambda\nu}p_{3\lambda}(p_2 -p_3)_{\beta}, 
\nonumber \\
\mathcal{M}_{8e}^{\mu\nu\alpha}&=&\frac{gg_{\rho\rho\eta}}{2\sqrt{6}}
\frac{1}{\sqrt{(1+ \frac{2}{3}c_A)}}\nonumber \\
&&\times[-(p_3+2p_4)^\nu g^{\gamma\alpha}-(p_3 -p_4)^\gamma
g^{\nu\alpha}+(2p_3+p_4)^\alpha g^{\nu\gamma}]  \nonumber \\
&&\times\frac{1}{s-m^2_\rho} \left[-g_{\gamma\gamma^{\prime}}+\frac{(p_1
+p_2)_\gamma (p_1 +p_2)_{\gamma^{\prime}}}{m_\rho^2}\right]
\epsilon^{\beta\mu\lambda\gamma^{\prime}}p_{1\beta}(p_1 +p_2)_\lambda.
\end{eqnarray}

\subsection{$\protect\rho\protect\eta\to\protect\rho\protect\rho$}

The amplitude for this reaction is given by 
\begin{eqnarray}
\mathcal{M}_{\rho\eta\to\rho\rho}=i\epsilon_{abc}(\mathcal{M}%
_{9a}^{\mu\nu\alpha} +\mathcal{M}_{9b}^{\mu\nu\alpha}+\mathcal{M}%
_{9c}^{\mu\nu\alpha}) \epsilon_{1\mu}\epsilon_{3\nu}\epsilon_{4\alpha},
\end{eqnarray}
with 
\begin{eqnarray}
\mathcal{M}_{9a}^{\mu\nu\alpha}&=& \frac{gg_{\rho\rho\eta}}{\sqrt{6}}
\frac{1} {\sqrt{1+\frac{2}{3}c_A}}\nonumber \\
&&\times[(2p_1-p_3)^\nu g^{\mu\gamma}+(2p_3 -p_1)^\mu g^{\gamma\nu}
-(p_1+p_3)^\gamma g^{\mu\nu}]  \nonumber \\
&&\times\frac{1}{t-m^2_\rho}\left[-g_{\gamma\gamma^{\prime}} +\frac{(p_1
-p_3)_\gamma(p_1 -p_3)_{\gamma^{\prime}}}{m_\rho^2}\right]
\epsilon^{\beta\gamma^{\prime}\lambda\alpha}p_{4\lambda}(p_2 -p_4)_{\beta}, 
\nonumber \\
\mathcal{M}_{9b}^{\mu\nu\alpha}&=&\frac{gg_{\rho\rho\eta}}{\sqrt{6}}
\frac{1} {\sqrt{1+\frac{2}{3}c_A}}  \nonumber \\
&&\times[(2p_1-p_4)^\alpha g^{\mu\gamma}+(2p_4 -p_1)^\mu
g^{\gamma\alpha}-(p_1+p_4)^\gamma g^{\mu\alpha}]  \nonumber \\
&&\times\frac{1}{u-m^2_\rho}\left[-g_{\gamma\gamma^{\prime}} +\frac{(p_1
-p_4)_\gamma(p_1 -p_4)_{\gamma^{\prime}}}{m_\rho^2}\right]
\epsilon^{\beta\gamma^{\prime}\lambda\nu}p_{3\lambda}(p_2 -p_3)_{\beta}, 
\nonumber \\
\mathcal{M}_{9c}&=& \frac{gg_{\rho\rho\eta}}{\sqrt{6}}
\frac{1} {\sqrt{1+\frac{2}{3}c_A}}\nonumber \\
&&\times[-(p_3+2p_4)^\nu g^{\gamma\alpha}-(p_3 -p_4)^\gamma g^{\nu\alpha}
+(2p_3+p_4)^\alpha g^{\nu\gamma}]  \nonumber \\
&&\times\frac{1}{s-m^2_\rho} \left[-g_{\gamma\gamma^{\prime}} +\frac{(p_1
+p_2)_\gamma(p_1 +p_2)_{\gamma^{\prime}}}{m_\rho^2}\right]
\epsilon^{\beta\mu\lambda\gamma^{\prime}}p_{1\beta}(p_1 +p_2)_\lambda.
\end{eqnarray}

\subsection{$\protect\rho\protect\eta\to\protect\pi\protect\omega$ and $%
\protect\rho\protect\eta\to\protect\pi\protect\pi$}

The amplitudes $\mathcal{M}_{\rho\eta\to\pi\omega}$ for the reaction $%
\rho\eta\to\pi\omega$ and $\mathcal{M}_{\rho\eta\to\pi\pi}$ for the reaction 
$\rho\eta\to\pi\pi$ can be obtained from the amplitudes for the reaction $%
\pi\eta\to\rho\omega$ and $\pi\eta\to\pi\rho$ via crossing symmetry, i.e.,
interchanging $p_1$ with either $-p_3$ or $-p_4$. Explicitly, they are given
by 
\begin{eqnarray}
\mathcal{M}_{\rho\eta\to\pi\omega}&=&\mathcal{M}_{\pi\eta\to\rho\omega}
(p_1\leftrightarrow -p_3),  \nonumber \\
\mathcal{M}_{\rho\eta\to\pi\pi}&=&\mathcal{M}_{\pi\eta\to\pi\rho}
(p_1\leftrightarrow -p_4).
\end{eqnarray}

\section{eta absorption by omega meson}

\label{omega}

The amplitudes for these reactions can be obtained from corresponding ones
for eta absorption by $\rho^0$ meson by replacing the mass of the $\rho$
meson with that of $\omega$ meson or by the crossing symmetry via
interchanging $p_1$ with $-p_4$, i.e., 
\begin{eqnarray}
\mathcal{M}_{\omega\eta\to K\bar K}&=&\mathcal{M}_{\rho^0\eta\to K\bar K}
(m_\rho\to m_\omega),  \nonumber \\
\mathcal{M}_{\omega\eta\to K \bar K^*(\bar KK^*)}&=& \mathcal{M}%
_{\rho^0\eta\to K \bar K^*(\bar KK^*)} (m_\rho\to m_\omega),  \nonumber \\
\mathcal{M}_{\omega\eta\to K^*\bar K^*}&=&\mathcal{M}_{\rho^0\eta\to K^*\bar
K^*} (m_\rho\to m_\omega),  \nonumber \\
\mathcal{M}_{\omega\eta\to\pi\rho}&=&\mathcal{M}_{\rho\eta\to\pi\omega}
(p_1\leftrightarrow -p_4).
\end{eqnarray}

\section{eta absorption by $K$ meson}

\label{kaon}

The amplitudes for the final states of $\pi K$, $\pi K^*$, $\rho K$, and $%
\rho K^*$ can be obtained from those for the reactions shown in Figs. \ref%
{dpion} and \ref{drho} using the crossing symmetry. For the final states of $%
\omega K$ and $\omega K^*$, the amplitudes are related to those with final
states $\rho^0 K$ and $\rho^0 K^*$ by replacing the mass of $\rho$ meson with
that of $\omega$ meson, i.e., 
\begin{eqnarray}
\mathcal{M}_{K\eta\to\pi K}&=&\mathcal{M}_{\pi\eta\to K\bar K}
(p_1\leftrightarrow -p_3),  \nonumber \\
\mathcal{M}_{K\eta\to\pi K^*}&=&\mathcal{M}_{\pi\eta\to K\bar K^*}
(p_1\leftrightarrow -p_3),  \nonumber \\
\mathcal{M}_{K\eta\to\rho K}&=&\mathcal{M}_{\rho\eta\to K\bar K}
(p_1\leftrightarrow -p_3),  \nonumber \\
\mathcal{M}_{K\eta\to\rho K^*}&=&\mathcal{M}_{\rho\eta\to K\bar K^*}
(p_1\leftrightarrow -p_3),  \nonumber \\
\mathcal{M}_{K\eta\to\omega K}&=&\mathcal{M}_{K\eta\to\rho^0 K} (m_\rho\to
m_\omega),  \nonumber \\
\mathcal{M}_{K\eta\to\omega K^*}&=&\mathcal{M}_{K\eta\to\rho^0 K^*}
(m_\rho\to m_\omega).
\end{eqnarray}

\section{eta absorption by $K^*$ meson}

\label{kstar}

Their amplitudes can also be obtained from those for other reactions via the
crossing symmetry or replacing the mass of $\rho$ meson with that of $\omega$
meson, i.e., 
\begin{eqnarray}
\mathcal{M}_{K^*\eta\to\pi K}&=&\mathcal{M}_{K\eta\to\pi K^*}
(p_1\leftrightarrow -p_3),  \nonumber \\
\mathcal{M}_{K^*\eta\to\pi K^*}&=&\mathcal{M}_{\pi\eta\to K^*\bar K^*}
(p_1\leftrightarrow -p_4),  \nonumber \\
\mathcal{M}_{K^*\eta\to\rho K}&=&\mathcal{M}_{K\eta\to\rho K^*}
(p_1\leftrightarrow -p_4),  \nonumber \\
\mathcal{M}_{K^*\eta\to\rho K^*}&=&\mathcal{M}_{\rho\eta\to K^*\bar K^*}
(p_1\leftrightarrow -p_3),  \nonumber \\
\mathcal{M}_{K^*\eta\to\omega K}&=&\mathcal{M}_{K^*\eta\to\rho^0 K}
(m_\rho\to m_\omega),  \nonumber \\
\mathcal{M}_{K^*\eta\to\omega K^*}&=&\mathcal{M}_{K^*\eta\to\rho^0 K^*}
(m_\rho\to m_\omega).
\end{eqnarray}


\begin{thebibliography}{99}
\bibitem{eta} O. Schwalb \textit{et al.}, Phys. Lett. B \textbf{321}, 20
(1994); F.D. Bert \textit{et al.}, Phys. Rev. Lett. \textbf{72}, 977 (1994);
R. Averbeck \textit{et al.}, Z. Phys. A \textbf{359}, 65 (1997).

\bibitem{dilepton} G. Q. Li, C. M. Ko, and G. E. Brown, Phys. Rev. Lett. 
\textbf{75}, 4007 (1995).

\bibitem{lin} Z. W. Lin, C. M. Ko, and S. Pal, Phys. Rev. Lett. \textbf{89},
152301 (2002).

\bibitem{n1535} A. De Paoli, K. W. Cassing, U. Mosel, and C. M. Ko, Phys.
Lett. B \textbf{219}, 194 (1989).

\bibitem{li} G. Q. Li, C. M. Ko, and G. E. Brown, Nucl. Phys. A \textbf{606}%
, 568 (1996); G. Q. Li, C. M. Ko, G. E. Brown, and H. Sorge, \textit{ibid.}, 
\textbf{611}, 539 (1996).

\bibitem{bando} M. Bando, T. Kugo, S. Uehara, K. Yamawaki, and T. Yanagida,
Phys. Rev. Lett. \textbf{54}, 1215 (1985).

\bibitem{song} C. Song, S. H. Lee, and C. M. Ko, Phys. Rev. C \textbf{52},
R476 (1995): C. Song, V. Koch, S. H. Lee, and C. M. Ko, Phys. Lett. B 
\textbf{366}, 379 (1996).

\bibitem{koch} L. Alvarez-Ruso and V. Koch, Phys. Rev. C \textbf{65}, 054901
(2002).

\bibitem{black} D. Black, A. H. Fariborz, and J. Schechter,Phys. Rev. D 
\textbf{61}, 074030 (2000)

\bibitem{ksfr} K. Kawarabayashi and M. Suzuki, Phys. Rev. Lett. {\bf 16},
255 (1966); Riazuddin and Fayyazuddin, Phys. Rev. {\bf 147}, 1071 (19660.

\bibitem{sakurai} J. Sakurai, Ann. Phys. {\bf 11}, 1 (1960). 

\bibitem{bramon74} A. Bramon, Phys. Lett. B {\bf 51}, 87 (1974).

\bibitem{particle} D. E. Groom {\it et al.}, Particle Data Group, Eur. 
Phys. J. C {\bf 15}, 1 (2000).

\bibitem{bramon} A. Bramon, A. Grau, and G. Pancheri, Phys. Lett. B \textbf{%
345}, 263 (1995).

\bibitem{fujiwara} T. Fujiwara, T. Kugo, H. Terao, S. Uehara, and K.
Yamawaki, Prog. Theor. Phys. \textbf{73}, 926 (1985).

\bibitem{jpsi1} Z. W. Lin and C. M. Ko, Phys. Rev. C \textbf{62}, 034903
(2000).

\bibitem{jpsi2} W. Liu, C. M. Ko, and Z. W. Lin, Phys. Rev. C \textbf{65},
015203 (2001).

\bibitem{charm} Z. W. Lin, T. G. Di, and C. M. Ko, Nucl. Phys. A \textbf{689}%
, 965 (2001).

\bibitem{liu} W. Liu, S. H. Lee, and C. M. Ko, Nucl. Phys. A \textbf{724},
375 (2003); W. Liu, C. M. Ko, and S. H. Lee, \textit{ibid.} \textbf{728},
457 (2003).

\bibitem{liu1} W. Liu and C. M. Ko, Phys. Rev. C \textbf{68}, 045203 (2003).

\bibitem{chli} C. H. Li and C. M. Ko, Nucl. Phys. A \textbf{712}, 110 (2002).

\bibitem{form} J. Vermaseren, computer code FORM, 1989. Free version of the
software is available on the Internet at
ftp://hep.itp.tuwien.ac.at/pub/Form/PC/.

\bibitem{xia} L. H. Xia and C. M. Ko, Phys. Rev. C \textbf{38}, 179 (1988);
L. H. Xia, C. M. Ko, and C. T. Li, Phys. Rev. C \textbf{41}, 572 (1990); G.
E. Brown, C. M. Ko, Z. G. Wu, and L. H. Xia, Phys. Rev. C \textbf{43}, 1881
(1991).

\bibitem{chen} L. W. Chen, V. Greco, C. M. Ko, S. H. Lee, and W. Liu, Phys.
Lett. B \textbf{601}, 34 (2004).

\bibitem{braun} P. Braun-Munzinger, D. Majestra, K. Redlich, and J. Stachel,
Phys. Lett. B \textbf{518}, 41 (2001).

\bibitem{greco} V. Greco, C. M. Ko, and P. L\'evai, Phys. Rev. Lett. \textbf{%
90}, 202302 (2003); Phys. Rev. C \textbf{68}, 034904 (2003).
\end{thebibliography}
\end{document}